
\input phyzzx
\overfullrule=0pt
\def\inbar{\,\vrule height1.5ex width.4pt depth0pt}
\def\IC{\relax{\hbox{$\inbar\kern-.3em{\rm C}$}}}
\def\IR{\relax{\rm I\kern-.18em R}}
\font\cmss=cmss10 \font\cmsss=cmss10 at 7pt
\def\IZ{\relax\ifmmode\mathchoice
{\hbox{\cmss Z\kern-.4em Z}}{\hbox{\cmss Z\kern-.4em Z}}
{\lower.9pt\hbox{\cmsss Z\kern-.4em Z}}
{\lower1.2pt\hbox{\cmsss Z\kern-.4em Z}}\else{\cmss Z\kern-.4em Z}\fi}
\tolerance=5000
\overfullrule=0pt
\twelvepoint
\def\comp{\circ}
\def\d{{\rm d}}

\singlespace
\pubnum{IASSNS-HEP-92/15}
\date{March, 1992}
\titlepage
\title{TWO DIMENSIONAL GAUGE THEORIES REVISITED}
\vglue-.25in
\author{Edward Witten\foot{Research supported in part by NSF Grant
PHY91-06210.}}
\medskip
\address{School of Natural Sciences
\break Institute for Advanced Study
\break Olden Lane
\break Princeton, NJ 08540}
\bigskip
\abstract{Two dimensional quantum Yang-Mills theory is reexamined
using a non-abelian version of the Duistermaat-Heckman integration
formula to carry out the functional integral.
This makes it possible to explain properties of the theory
that are inaccessible to standard methods and to obtain
general expressions for intersection pairings on moduli spaces of
flat connections on a two dimensional surface.  The latter expressions
agree, for gauge group $SU(2)$, with formulas obtained recently
by several methods.

}

\endpage
\def\ZZ{\widetilde Z}
\chapter{Introduction}
\REF\migdal{A. Migdal, Zh. Eksp. Teor. Fiz. {\bf 69} (1975) 810
(Sov. Phys. Jetp. {\bf 42} 413).}
\REF\bralic{N. Bralic, Phys. Rev. {\bf D22} (1980) 3090.}
\REF\kaz{V. Kazakov and I. Kostov, Nucl. Phys. {\bf B176} (1980) 199,
Phys. Lett. {\bf B105} (1981) 453, Nucl. Phys. {\bf B179} (1981) 283.}
\REF\gks{L. Gross, C. King, and A. Sengupta, Ann. of Phys. {\bf 194} (1989)
65.}
\REF\migstud{B. Rusakov, ``Loop Averages And Partition Functions in $U(N)$
Gauge Theory On Two-Dimensional Manifolds,'' Mod. Phys. Lett. {\bf A5}
(1990) 693.}
\REF\fine{D. Fine, ``Quantum Yang-Mills On The Two-Sphere,'' Commun. Math.
Phys. {\bf 134} (1990) 273, ``Quantum Yang-Mills On A Riemann Surface,''
to appear in Commun. Math. Phys.}
\REF\seng{A. Sengupta, ``The Yang-Mills Measure For $S^2$,'' to appear in
J. Funct. Anal., ``Quantum Gauge Theory On Compact Surfaces,'' preprint.}
\REF\blau{M. Blau and G. Thompson, ``Quantum Yang-Mills Theory On
Arbitrary Surfaces,'' preprint NIKHEF-H/91-09,
MZ-TH/91-17.}
\REF\ewitten{E. Witten, ``On Quantum Gauge Theories In Two Dimensions,''
Commun. Math. Phys. {\bf 141} (1991) 153.}
\REF\forman{R. Forman, ``Small Volume Limits of $2-d$ Yang-Mills,'' preprint,
Rice University (1991).}
\REF\uwitten{E. Witten, ``On The Structure Of The Topological Phase
Of Two Dimensional Gravity,'' Nucl. Phys. {\bf B340} (1990) 281.}
\REF\distler{J. Distler, ``$2-D$ Quantum Gravity, Topological Field Theory,
and the Multicritical Matrix Models,'' Nucl. Phys. {\bf B342} (1990) 523.}
\REF\kontsevich{M. Kontsevich, ``Intersection Theory On The Moduli Space
Of Curves And The Matrix Airy Function,'' preprint (1991).}
\REF\schwarz{A. Schwarz, ``The Partition Function Of Degenerate Quadratic
Functional And Ray-Singer Invariants,'' Lett. Math. Phys. {\bf 2} (1978) 247.}
\REF\duistermaat{J. J. Duistermaat and G. J. Heckman, ``On The Variation
In The Cohomology In The Symplectic Form Of The Reduced Phase Space,''
Invent. Math. {\bf 69} (1982) 259.}
\REF\bismut{J.-M. Bismut, ``Localization Formulas, Superconnections,
and The Index Theorem For Families,'' Comm. Math. Phys. {\bf 103} (1986) 127.}
\REF\abot{M. F. Atiyah and R. Bott, ``The Moment Map And Equivariant
Cohomology,'' Topology {\bf 23} (1984) 1.}
\REF\abott{M. F. Atiyah and R. Bott, ``The Yang-Mills Equations Over Riemann
Surfaces,'' Phil. Trans. Roy. Soc. London {\bf A308} (1982) 523.}
\REF\kirwan{F. Kirwan, {\it Cohomology of Quotients In Symplectic
And Algebraic Geometry,} Princeton University Press.}

This paper will be devoted to a renewed study of two dimensional
Yang-Mills theory without matter, a system
which can be easily solved [\migdal] and has been extensively studied
[\bralic--\forman].  Yet we will see that there is still much to say
about this supposedly ``trivial'' system.  To state our result in a nutshell,
we will explain (in \S3)
a simple mapping from topological Yang-Mills
theory to physical Yang-Mills theory in two dimensions -- analogous
to the far more mysterious equivalence of topological and physical
gravity in two dimensions [\uwitten--\kontsevich].

What can be learned from this?  To begin with the physics,
as in the case of any gauge theory, one can attempt to
expand the partition function $Z(\epsilon)$
of two dimensional Yang-Mills theory in powers of the gauge
coupling constant $\epsilon$.  In doing so (in a suitable topological
sector), one finds a remarkable result: the perturbation series in
$\epsilon$ stops after finitely many terms, yet $Z(\epsilon)$ is not
a polynomial.  $Z(\epsilon)$ contains exponentially small terms which
can be identified as contributions of {\it unstable} classical solutions
to the functional integral.  (A solution with $n$ unstable modes
is weighted with a phase of $i^n$ -- it turns out that $n$ is always even.)
Conventional physical methods are quite inadequate for explaining
such behavior.  It turns out that this can be done using the relation
that we will find between physical and topological gauge theories or
differently put using
a generalization to problems with non-abelian symmetries
of the exact integration formula of Duistermaat and Heckman
[\duistermaat].  We will devote \S2-3 to an explanation of the necessary
ideas, first from a mathematical standpoint in \S2, and then more physically
in \S3.
The argument in \S2 uses an idea similar to that in a proof by
Bismut [\bismut] of the DH formula.
In \S4, we will apply our integration formula to two dimensional
gauge theories, explaining the peculiar properties of the function
$Z(\epsilon)$.

\REF\guillemin{V. Guillemin and E. Prato, ``Heckman, Kostant, and Steinberg
Formulas for Symplectic Manifolds,'' Adv. Math. {\bf 82} (1990) 160.}
The usual Duistermaat-Heckman (DH) formula can be applied to problems with
non-abelian group action [\guillemin], and it may well be that in finite
dimensions, most of the applications of our formula
can be deduced from the DH formula.
Even if this is so,
the formulation we give is natural in infinite dimensions, as
should be clear in \S3-4.

Mathematically, it has been known [\blau,\ewitten,\forman] that $Z(\epsilon)$
can be expressed at $\epsilon=0$
in terms of the volumes of the moduli spaces
of flat connections on a surface; this is essentially a consequence
of old work by A. Schwarz [\schwarz] and was used in [\ewitten] to obtain
precise formulas for these volumes.  Yet a study of the function $Z(\epsilon)$
has suggested that the relation of this function to the topology of the moduli
spaces is not limited to $\epsilon=0$.  Coming to grips with this phenomenon
is the goal of the present paper from a mathematical standpoint.
This
requires the nonabelian integration formulas of \S2-3 and their application
to infinite dimensional functional integrals in \S4-5.
The upshot will be precise and general formulas for intersection pairings
on moduli spaces of flat connection, which are presented in \S5.

Readers of this paper may want to bear in mind that \S2 and \S3 cover
much the same ground and are independent to a large extent.  Some readers
might prefer to focus on \S3 after just glancing at \S2.

We now turn to a more extensive introduction
to our subject.

\subsection{Integration Formulas}

The DH formula, which has many
fascinating applications,
governs the following situation.  Let $X$ be a $2n$ dimensional
compact symplectic manifold,
with symplectic form $\omega$.  Suppose that the group $U(1)$ acts
symplectically
on $X$, the action being
generated by a vector field $V$.  The action is said to be Hamiltonian
if there is a function $H$ on $X$ such that  $\d H=-j_V\omega$.
\foot{I will use the symbol $j_V$ to denote contraction with a vector field
$V$, so
in local coordinates $x^i$, if $\omega={1\over 2}\omega_{ij}dx^i\wedge dx^j$,
then  $j_V\omega=V^i\omega_{ij}dx^j$. This operator
is more usually written as $i_V$, but I want to avoid confusion with
$i=\sqrt{-1}$.  For a one form $\lambda $, I also write $j_V(\lambda)$ as
$\lambda(V)$.}
The partition
function of classical statistical mechanics, with this phase space and
Hamiltonian, is
$$ \int_X{\omega^n\over n!} e^{-\beta H}. \eqn\pilm$$
The DH formula asserts that this integral is given
exactly by the semi-classical
approximation, provided that one sums over all critical
points of $H$.  If the critical points are isolated points $P_i$, then
the formula is
$$ \int_X{\omega^n\over n!} e^{-\beta H}=
\sum_i {e^{-\beta H(P_i)} \over \beta^n e(P_i)} , \eqn\ilm$$
where $e(P_i)$ is the product of the weights of the circle action in the
tangent space at $P_i$; this factor can be interpreted as the determinant
arising from a Gaussian integral near $P_i$. (The sign of $e(P_i)$ is
$(-1)^{n_i/2}$, where $n_i$, which is even because of the circle action,
is the Morse index of $P_i$ -- as if each unstable mode contributes a factor
of $i$ to the integral.)

The DH formula is usually stated for oscillatory integrals
-- imaginary $\beta$ -- in which case \ilm\ is the assertion of exactness
of the
stationary phase approximation (summed over critical points).
In our applications, real $\beta$ is more natural.
A simple example of the application of the
DH formula is given in the appendix.

The DH formula has a cohomological interpretation
[\abot] which shows that the basic principle is
not stationary phase
but localization at the fixed points of the $U(1)$ action; in this form
\ilm\ can be generalized to a larger class of integrals.

\subsection{The Non-Abelian Case}

Suppose we are given the action on $X$ not of $U(1)$ but of a compact,
connected Lie group $G$, with Lie algebra ${\cal G}$.
The action of $G$ is said to be Hamiltonian if it is induced from
a homomorphism $\widetilde\mu:{\cal G}\to {\rm Fun}(X)$, where ${\rm Fun}(X)$
is
the space of smooth
functions on $X$, regarded as a Lie algebra via the Poisson bracket.
This amounts to saying that for every element $T_a$ of ${\cal G}$,
represented by a vector field $V_a$ on $X$, there is a corresponding
Hamiltonian function $\mu_a$, with $j_{V_a}(\omega)=-\d\mu_a$, and with
the map $T_a\to \mu_a$ being a homomorphism.
The $\mu_a$ can be assembled into a map $\mu:X\to {\cal G}^*$ (${\cal G}^*$ is
the dual of ${\cal G}$).

In particular, we do not have a single
$G$ invariant function $\mu$, but a collection of them; so
we must modify \pilm.
To this end, we introduce an invariant quadratic
form $(~,~)$ on ${\cal G}$ and consider the integral
$$Z =\int_M{\omega^n\over n!}\,\,\exp\left({-{\beta\over 2}(\mu,\mu)}\right)
. \eqn\juju$$
The critical point set of the function $I=(\mu,\mu)$ that appears here
is very special, since
according to Atiyah and Bott [\abott] and Kirwan [\kirwan], $I$
is an equivariantly perfect Morse function.
In this paper, we will use the critical points of this function in another
way.
We will see that there is
an analog of the
DH formula expressing $Z$ as a sum of contributions
of critical points of $I$.
The general statement is of the following form.  Let $S$ be the set of
components of the critical point set of $I$.  For every component
$X_\alpha$ of the critical point set, there is a function $Z_\alpha(\beta)$,
determined by the local behavior of $\omega$ and $\mu$
near $X_\alpha$ up to some finite order,
such that
$$Z(\beta)=\sum_{\alpha\in S} Z_\alpha(\beta). \eqn\offc$$
The $Z_\alpha(\beta)$ can be very complicated functions of $\beta$.
Transcendental
functions such as the error function arise even in the simple abelian
example
treated in the appendix.

However, a simple contribution
arises in one important special case, which will be the
basis for our applications.  The absolute minimum of $I$ -- which will give
the dominant contribution in the important limit of $\beta\to\infty$ --
is $\mu^{-1}(0)$.  The quotient
${\cal M}=\mu^{-1}(0)/G$ is called the reduced phase space or symplectic
quotient of $X$ by $G$.  $\mu^{-1}(0)/G$ is naturally a symplectic manifold
with a symplectic form that we will also call $\omega$.
If $\mu^{-1}(0)$ is a smooth manifold, on which $G$ acts freely,
then the contribution of $\mu^{-1}(0)$ to $Z$ is given by a simple
cohomological formula that we will explain, essentially
$${1\over \beta^{{1\over 2}{\rm dim}G}}\cdot
 \int_{{\cal M}}\exp\left({\omega +{1\over 2\beta}\Theta}\right)
    \eqn\gomo$$
where $\Theta$ is a certain
element of $H^4({\cal M},\IR)$ that will appear in due
course.  Thus, $Z$ differs from \gomo\ only by terms that vanish exponentially
for $\beta\to\infty$.

\gomo\ is analogous to a cohomological formula of DH
for the pushforward of $\omega^n/n!$ by the moment map.
Like the DH formula, our formula for $Z$ comes from a localization
principle which applies to a larger class of integrals; in fact we will
need the generalization.

\subsection{Application To Gauge Theories}

Now let us describe the infinite dimensional setting in which we will apply
these considerations.

Let $H$ be a compact, connected (but perhaps not simply connected)
Lie group with Lie algebra ${\cal H}$.
We will assume $H$ is simple.
The extension to
more general compact $H$ does not involve essentially new ideas.

For $H=SU(N)$, introduce a quadratic form $(~,~)$
on ${\cal H}$ by
$$(a,b)=-\Tr ab,     \eqn\modoc$$
where $\Tr $ is the trace in the $N$ dimensional representation.
This has the property that the fundamental integer-valued
characteristic number
of a rank $N$ vector bundle $E$ with structure group $SU(N)$ on a closed
four-manifold
$X$ is
$$- {1\over 8\pi^2}\int_X\Tr F\wedge F, \eqn\odoc$$
where $F$ is the curvature of a connection on $E$.  (This integrality
insures integrality of the symplectic form introduced presently.)
For any simple connected $H$, we define $-\Tr$ to be
a positive definite quadratic form on ${\cal H}$
such that \odoc\ is the fundamental characteristic number of an
$\widetilde H$
bundle over a a four-manifold, $\widetilde H$
being the universal cover of $H$.

\def\ade{{\rm ad}(E)}
Let $\Sigma$ be an oriented closed Riemann surface of genus $g$.
Let $E$ be an $H$ bundle over $\Sigma$ (one can think in terms of
a principal bundle or
a vector bundle with a reduction of the structure group to $H$).  The adjoint
vector bundle associated with $E$ will be called ${\rm ad}(E)$.
Let ${\cal A}$ be the space of connections on $E$.  The space of connections
is an affine space whose tangent space can be identified with
$\Omega^1(\Sigma,{\rm ad}(E))$ (that is, the space of ${\rm ad}
(E)$-valued one-forms
on $\Sigma$).  This being so, a symplectic form on ${\cal A}$
can be defined by
$$\omega(a,b)={1\over 4\pi^2}\int_\Sigma\Tr (a\wedge b).  \eqn\urgo$$

Let $G$ be the group of gauge transformations on $E$.
The Lie algebra ${\cal G}$ of $G$ is the space of ${\rm ad}(E)$-valued
zero-forms; the dual of the Lie algebra consists of ${\rm ad}(E)$-valued
two-forms.  $G$ acts symplectically on ${\cal A}$, with a moment
map given [\abott] by the map
$$\mu(A)=-{F\over 4\pi^2},  \eqn\hccx$$
from the connection $A$ to its ${\rm ad}(E)$-valued curvature
two-form $F=\d A+A\wedge A$.

$\mu^{-1}(0)$ therefore consists of flat connections, and $\mu^{-1}(0)/G$
is the moduli space ${\cal M}$ of flat connections on $E$ up to gauge
transformation.  ${\cal M}$ is a component of the moduli space of homomorphisms
$\rho:\pi_1(\Sigma)\to H$, up to conjugation.

Endow $\Sigma$ with a measure $\mu$ of total area 1.
This determines a metric or
quadratic form $(a,a)=-\int_\Sigma d\mu \cdot\Tr a^2$
on ${\cal G}$;  hence it determines on the dual of ${\cal G}$ a quadratic
form which we can write as $(F,F)=-\int_\Sigma d\mu \Tr f^2$ where
$f=\star F$.
(Here $\star$ is the Hodge star operator; we recall that in two dimensions
the $\star$ operator between two-forms and zero-forms depends only on
a measure, not a metric.)
The partition function of two dimensional quantum Yang-Mills theory
on the surface $\Sigma$ is formally given by the Feynman path integral
$$Z(\epsilon)={1\over {\rm vol}(G)}\int_{\cal A} DA\,\,\exp\left(
{-{1\over 2\epsilon}(F,F)}\right),
 \eqn\uurgo$$
where $\epsilon$ is a real constant,
$DA$ is the symplectic measure on the infinite dimensional
function space ${\cal A}$,
and ${\rm vol}(G)$ is the volume of $G$ (determined
formally from the volume form on $G$ associated with the metric on ${\cal G}$).

Local considerations due to Migdal
[\migdal] can be adapted in various ways
[\bralic--\forman]
to give a rather direct computation of \uurgo\ (and various closely related
integrals).
On the other hand, since $(F,F)$ is the norm of the moment map with respect
to an invariant metric on the Lie algebra, the integral
\uurgo\ is precisely of the form of the integrals \juju\
governed by the new localization principle that we will present.
Comparing the known $Z(\epsilon)$ to predictions of the localization principle,
we will find full agreement for all properties that can be computed on both
sides, including some surprising properties of $Z(\epsilon)$
that we mentioned at the
outset.

\subsection{The Cohomology Ring Of The Moduli Space}

Mathematically,
the payoff comes by looking closely at the contribution of $\mu^{-1}(0)$
to the critical point formula \offc.
This contribution can be extracted
from the small $\epsilon$ behavior of $Z(\epsilon)$;
on the other hand, it has an interpretation
in terms of the topology of ${\cal M}$ that emerges in the proof  of
the critical point formula.  Comparing these will give our main
topological conclusions.

The contribution of $\mu^{-1}(0)$ to \offc\
will turn out to have a simple cohomological interpretation
when the gauge group and bundle are such that
$\mu^{-1}(0)$ is non-singular, and acted on freely by $G$.  In practice,
this occurs in genus $\geq 1$
for $H=SU(N)/\Gamma$, with $\Gamma$ a subgroup of the
center of $SU(N)$, and certain bundles $E$.  In the rest
of this introduction, we consider only such cases. However,
even when there are singularities,
the fixed point theorem gives in principle
a more sophisticated formula, deserving of study,
for the contribution of $\mu^{-1}(0)$.

\REF\newstead{P. E. Newstead, ``Characteristic  Classes Of Stable Bundles
Over An Algebraic Curve,'' Trans. Am. Math. Soc. {\bf 169} (1972) 337.}
Explicit
generators $x_i$ for $H^*({\cal M},\IR)$  are known
[\newstead,\abott].  We will recall their
definition in \S3.3.
By comparing the direct evaluation
of $Z(\epsilon)$ to the predictions of the critical point formula, we will
obtain in the nonsingular case
explicit expressions for all of the quantities
$$\int_{\cal M} x_1\wedge x_2\wedge \dots\wedge x_n. \eqn\ombo$$
This is tantamount to a determination of the cohomology
ring of ${\cal M}$.

\REF\thaddeus{M. Thaddeus,  ``Conformal Field Theory And The Cohomology Of
The Moduli Space Of Stable Bundles,'' Oxford preprint (1991).}
\REF\fkirwan{F. Kirwan, ``The Cohomology Rings Of Moduli Spaces Of
Vector Bundles Over Riemann Surfaces,'' Oxford preprint.}
\REF\donaldson{S. K. Donaldson, ``Gluing Techniques In The Cohomology Of
Moduli Spaces,'' Oxford preprint.}
\REF\everlinde{E. Verlinde, ``Fusion Rules And Modular Transformations
In 2d Conformal Field Theory,'' Nucl. Phys. {\bf B300} (1988) 360.}
\REF\tsuchiya{A. Tsuchiya, K. Ueno, and Y. Yamada, ``Conformal Field THeory On
Universal Family Of Stable Curves With Gauge Symmetries,'' Adv. Studies
in Pure Mathematics {\bf 19} (1989) 459.}
\REF\segal{G. Segal, ``Two Dimensional Conformal Field Theories And Modular
Functors,'' in {\it IXth International Conference On Mathematical Physics},
eds. B. Simon, A. Truman, and I. M. Davies (Adam Hilger, Bristol, 1989);
``Fusion Of Representations Of Loop Groups,''
unpublished manuscript.}
\REF\sz{A. Szenes, ``Hilbert Polynomials of Moduli Spaces of Rank 2
Vector Bundles, I'' (Harvard preprint); A. Bertram and A. Szenes,
``Hilbert Polynomials of Moduli Spaces of Rank 2 Vector Bundles, II''
(Harvard preprint). }
\REF\dask{G. Daskalopoulos and R. Wentworth, ``Geometric Quantization For The
Moduli Space Of Vector Bundles With Parabolic Structure,'' ``The Integrability
Criterion in the $SU(2)$ Chern-Simons Gauge Theory,'' MIT preprints.}
\REF\chang{S. Chang, ``Non-Abelian $\theta$-Functions, Geometric Quantization,
And A Mathematical Proof of Verlinde's Formula.''}
The intersection numbers \ombo\ have been the subject of recent papers
[\thaddeus--\donaldson]
giving a complete answer for $H=SO(3)$ and $E$
a bundle of non-zero $w_2$.  Here is a brief synopsis of the methods
and comparison to the approach here.

Thaddeus [\thaddeus] uses the Verlinde formula of
conformal field theory [\everlinde--\chang].
Since not all the $x_i$ appear in the Verlinde formula, Thaddeus
supplements this with a geometrical argument to eliminate the extra
classes in the case of $SO(3)$.
The analogous step in our calculation
is the integration over $\psi$ and change of variables to reduce (5.11)
to (5.14), thereby eliminating (for any $H$) those
$x_i$ that do not appear in the Verlinde formula.
It would be interesting to know how Thaddeus's geometrical argument
could be extended to other $H$.  As for the main part of Thaddeus's paper,
which is the use of the Verlinde formula, the analog of this in the present
paper is the explicit evaluation of the Yang-Mills path integral (5.14)
by a gluing method.  The origins of the Verlinde formula (and most approaches
to proving it, though [\sz] is an exception) involve analogous but
more sophisticated gluing arguments.

Kirwan [\fkirwan]  uses considerations
involving the geometry of the space of connections
starting with the fact that $I$ is an equivariantly perfect Morse function.
She proves for $SO(3)$ (\ie, rank two bundles of odd degree) the completeness
of the Mumford relations among products of the $x_i$;
modulo a knowledge of the volumes of the moduli spaces, this  is equivalent
to evaluation of \ombo.
Her considerations, which in principle apply for any $H$,
probably underlie our nonabelian localization formula.

Donaldson [\donaldson]
uses a topological gluing construction ingeniously extracted from
work on the
Verlinde formula.  This makes it possible to evaluate the pairings \ombo\
by pure differential topology.

\chapter{Non-Abelian Localization}
\REF\quillen{V. Mathai and D. Quillen, ``Superconnections, Thom Classes, and
Equivariant Differential Forms,'' Topology {\bf 25} (1986) 85.}

\section{Equivariant Integration}

In this section, I will explain the non-abelian localization principle.

Let $X$ be a compact closed manifold acted on by
a compact connected Lie group $G$, with Lie algebra ${\cal G}$.

\def\fun{{\rm Fun}({\cal G})}
We must recall the de Rham model for the $G$-equivariant cohomology
of $X$ (see [\abot, pp. 10-13] and [\quillen] for explanations),
which we will take with complex coefficients.
Let $\Omega^*(X)$ be the de Rham complex
of $X$. Let ${\rm Fun}({\cal G})$ be the algebra of polynomial functions on
${\cal G}$, graded so that an $n^{\hbox{\it th}}$
order homogeneous polynomial is considered to be of degree $2n$.
We will later consider various completions of ${\rm Fun}({\cal G})$.
The desired complex is $\Omega_G^*(X)=\left(\Omega^*(X)
\otimes \fun\right)^G$, where $^G$ denotes the $G$-invariant part.
(For a description of the same thing in the notation of physicists,
see the beginning of \S3.)  An element of this complex
is called an equivariant differential form.

The $G$ action on $X$ is determined by a homomorphism from ${\cal G}$
to the Lie algebra ${\rm Vect}(X)$ of vector fields on $X$.  Let us
denote the vector field on $X$ corresponding to $\phi\in {\cal G}$
as $V(\phi)$.
One endows $\Omega_G^*(X)$ with the differential\foot{The factor
of $i=\sqrt{-1}$
in the last term here  is usually omitted in mathematical papers,
and plays no essential role, since it can be removed by conjugation.
I include it so that later formulas will agree with standard physics
conventions.}
$$D = \d -ij_{V(\phi)}. \eqn\ucu$$
We have
$$D^2=-i{\cal L}_{V(\phi)}, \eqn\cuu$$
with ${\cal L}_{V(\phi)}=\d j_{V(\phi)}+j_{V(\phi)}\d$
the Lie derivative with respect
to $V_a$.  Thus, $D^2=0$ precisely on the $G$-invariant subspace
$\Omega_G^*(X)$
of $\Omega^*(X)\otimes \fun$.  The cohomology of the $D$ operator
is called the $G$-equivariant cohomology of $X$, $H_G^*(X)$.

Now we want to introduce a notion of integration of equivariant differential
forms.  The operation usually
considered is the pushforward $\Omega_G^*(X)\to \Omega_G^*({\rm pt})$
obtained by integration over $X$:
$$ \alpha \to \int_X \alpha.       \eqn\ucux$$
To show that this descends to a map $H_G^*(X)\to H_G^*({\rm pt})$, one
must show that one can integrate by parts:
$$ 0 = \int_X D\beta. \eqn\hucu$$
This is true since $\int_X\d\beta =0$ by ordinary integration by parts,
and $\int_X j_{V(\phi)}\beta=0$ since $j_{V(\phi)}\beta$ does not have
a component which is a differential form of top dimension.

This, however, is not quite the integration operation that we want.
As a vector
space, ${\cal G}$ has a natural translation invariant measure, unique
up to a constant factor.  To fix that factor, note that as
${\cal G}$ is naturally isomorphic to the tangent space to $G$ at
the identity, a choice of Haar measure on $G$ determines a measure
on ${\cal G}$.  Picking an arbitrary Haar measure on $G$, with total
volume ${\rm vol}(G)$, let $\phi_1, \phi_2,
\dots ,\phi_s$ be Euclidean coordinates on ${\cal G}$ such that
the measure $\d\phi_1\d\phi_2\dots \d\phi_s$ on ${\cal G}$ coincides with the
chosen Haar measure at the identity of $G$. Then
$$ {1\over {\rm vol(G)}}\d\phi_1\d\phi_2\dots \d\phi_s \eqn\ommo$$
is a natural measure on ${\cal G}$, independent of the chosen Haar measure
on $G$.

The integration operation that we want is now roughly speaking the map
$\Omega_G^*(X)\to \IC$ given by
$$\alpha \to {1\over {\rm vol}(G)}\int_{{\cal G}\times X}{\d\phi_1\d\phi_2
\dots \d\phi_s\over (2\pi)^s}\cdot \alpha.   \eqn\mmo$$
This definition is, however, unsatisfactory as the integral does not
generally
converge.  If we complete $\fun$ to permit functions that are not necessarily
polynomials in $\phi$,
then \mmo\ converges for a suitable class of $\alpha$'s.
Since we want to allow a somewhat larger class, we introduce a convergence
factor in the definition.  Let $\epsilon$ be a positive real number.
Let $(~,~)$ be the positive definite invariant quadratic form on ${\cal G}$
described in the introduction.  The definition that we want is then
$$\oint \alpha ={1\over {\rm vol}(G)}\int_{{\cal G}\times X}
{\d\phi_1\dots \d\phi_s\over (2\pi)^s}\,\alpha\cdot
\exp\left({-{\epsilon\over 2}
(\phi,\phi)}\right).\eqn\dodo$$
This of course converges for forms with polynomial dependence on $\phi$.
Later, we will work not just with polynomial forms but with forms that
are permitted to have an exponential growth for large $\phi$;
\dodo\ also converges in this larger class.
We will call the operation in \dodo\ equivariant integration.

The same argument as above shows that $\oint D\beta=0$ for any $\beta$,
so $\oint$ descends to a map from $H_G^*(X)$ to $\IC$.  In fact,
this map is just the composition of ordinary integration
over $X$ with the map $H_G^*({\rm pt})\to \IC$ given by
$$ \gamma\to {1\over {\rm vol}(G)}\int {\d \phi_1\d \phi_2\dots \d
\phi_s\over (2\pi)^s}
\cdot \gamma.          \eqn\gopher$$
By mapping from $H_G^*({\rm pt})$
to $\IC$ we have, in a sense, discarded most
of the information.  The discarded information can be recovered
by considering not just $\oint\alpha$ but also
$\oint \alpha\cdot Q(\phi)$ with $Q$ an arbitrary $G$ invariant
polynomial on ${\cal G}$ (that is, an arbitrary element of
$\Omega_G^*({\rm pt})$).  Happily, the localization principle to which
we presently turn applies to all of these integrals.

\subsection{$G$ Action On A Point}

In general, the $\oint$ operation diverges as
$\epsilon\to 0$.  For instance, in the basic case $X={\rm a~point}$,
$\alpha = 1$, we have
$$\oint_{{\rm pt}} 1 = {1\over {\rm vol}(G)\cdot (2\pi\epsilon)^{s/2}}.
\eqn\pointtarget$$
This is an important formula which is the simplest illustration of
the relation of singularities of the $G$ action to singularities of
equivariant integrals as functions of $\epsilon$.

\section{The Localization Principle}

Now we will explain the localization principle.
Let $\alpha$ be any equivariantly closed form.  Then for any real
number $t$ and any
$\lambda\in \Omega_G^*(X)$ whose dependence on $\phi$ is sufficiently
mild,
$$\oint_X\alpha =\oint_X \alpha \,e^{tD\lambda}, \eqn\moxie$$
since $\alpha\left(1-e^{tD\lambda}
\right)$ can be written as $D\left(\dots\right)$ using $D\alpha=D^2=0$.
A restriction on the $\phi$ dependence of $\lambda$ must be imposed here
to ensure the convergence of the right hand side of \moxie\ and
to justify the integration by parts that is involved in proving \moxie.
We will consider only the case that $\lambda$ is independent of $\phi$.
In fact, we will suppose that $\lambda$ is a $G$-invariant one-form.

Pick an orthonormal basis $T_a$ of ${\cal G}$, and write
$V(\phi)=\sum_a\phi^aV_a$, where $V_a$ is the vector field on $M$
representing $T_a$, and the $\phi^a$ are linear functions on ${\cal G}$.
Then \moxie\ can be written out explicitly as
$$\oint_X\alpha ={1\over {\rm vol}(G)}\int {\d\phi_1\d\phi_2\dots\d\phi_s
\over (2\pi)^s}
\alpha\cdot \exp\left({t\,\,d\lambda-it\sum_a\phi^a\lambda(V_a)}
-{\epsilon\over 2}\sum_a(\phi^a)^2\right) . \eqn\murmor$$
If we suppose that $\alpha$ is independent of $\phi$, then we can perform
the Gaussian $\phi$ integral, to get
$$\oint_X\alpha={1\over {\rm vol}(G)\cdot (2\pi\epsilon)^{s/2}}
 \int_X \alpha\cdot \exp\left({t\,\,\d\lambda}
-{t^2\over 2\epsilon}\sum_a
\left(\lambda(V_a)\right)^2\right).     \eqn\urmor$$

\REF\arnold{V. I. Arnold, S. M. Gusein-Zade, and A. N. Varchenko,
{\it Singularities Of Differentiable Maps}, Vol. II (Birkhauser, Boston,
1988).}
The crucial factor in \urmor\ is the last exponential factor.
Let $X'$ be the subspace of $X$ on which
$$ \lambda(V_a)=0,~~~a=1,\dots, s. \eqn\mirrp$$
Write
$$X' = \cup_{\sigma\in S} X_\sigma, \eqn\irrp$$
where $X_\sigma$ are the connected components of $X'$, and $S$ is the
set of such components.  Let $W$ be a compact subset of $X$ with
$W\cap X'=\emptyset$.
We know that \urmor\ is independent of $t$.  On the other hand,
the integral over $W$ of the right hand side of \urmor\ vanishes
for $t\to \infty$ as $\exp(-ct^2)$ for some positive constant $c$.
Let $Z_\sigma$ be the integral of the right hand side of \urmor\ over a
tubular neighborhood of $X_\sigma$, in the limit $t\to\infty$.
$Z_\sigma$
is determined by the local behavior of $\alpha$ and the $G$ action near
$X_\sigma$ up to some finite order.  (In the case that $Z_\sigma$
is a point, this is a consequence of
theorem 7.6 of [\arnold].)  And taking the large $t$ limit of \urmor, we get
an expression
$$\oint_X\alpha=\sum_{\sigma\in S} Z_\sigma.     \eqn\umoc$$
for $\oint_X\alpha$ as a sum of local contributions.

So far we have assumed that $\alpha$ is independent of $\phi$.
It would not be different if $\alpha$ has a polynomial or even
exponential dependence on $\phi$.  The $\phi$ integral in \murmor\
would then contribute an extra polynomial or exponential $t$ dependence
on the right hand side of \urmor, too weak to affect the localization,
which was determined by a factor $\exp(-ct^2)$.  The detailed computation
of $Z_\sigma$ in the case that $\alpha$ has exponential $\phi$ dependence
requires care and will be considered later in a special situation.

\subsection{Stationary Phase}

The argument leading to localization can actually be formulated
without performing the $\phi$ integral.
Looking back to
\murmor, we see that apart from a polynomial in $t$ that comes
from the expansion of $\exp(t\,\,\d\lambda)$, the $t$ dependence
appears entirely in a factor $\exp(-itK)$, where $K$ is the function
$$K = \sum_a\phi^a \lambda(V_a) \eqn\mcco$$
on $X\times {\cal G}$.  The method of stationary phase (expounded, for
instance, in [\arnold], chapter 7)
enables one to compute the large $t$ behavior
of such an integral in terms of local data
on the critical point set of
$K$.  (The convergence factor $\exp(-{\epsilon\over 2}(\phi,\phi))
$ in our integral ensures that there are no essential problems
coming from the lack of compactness of $X\times {\cal G}$ and the possible
lack of compactness of the critical point set.)

The critical point condition $\d K=0$ gives two conditions. Varying
with respect to $\phi$, we get the familiar equation
$$ \lambda(V_a)=0, \,\,\,\,\,a=1\dots s. \eqn\hchc$$
Varying with respect to the coordinates of $X$, we get
$$\phi^a\,\,\d\left(\lambda(V_a)\right) = 0. \eqn\mcmc$$
The $\phi^a$ take values in a vector space, which is contractible
to the origin by scaling; and the equations are invariant under this
scaling.  So the homotopy type of the space of solutions would not be
changed if we restrict to $\phi=0$, and in particular
the components $\widetilde X_\sigma$ of the critical
set of $K$ are in one to one correspondence with
the components $X_\sigma$ of solutions of \hchc.  The contribution
of $\widetilde X_\sigma$ to the large $t$ evaluation of \murmor\
coincides with the contribution of $X_\sigma$ to the large $t$ evaluation
of \urmor, since it reduces to the latter upon performing the $\phi$ integral.

Because of the invariance under scaling of $\phi$,
$\widetilde X_\sigma$
is compact when and only when $\widetilde X_\sigma=X_\sigma$; this
is so precisely when \mcmc\ implies that $\phi^a=0$.
When this is so, compactness of $\widetilde X_\sigma$ means that
the convergence factor $\exp(-{\epsilon\over 2}(\phi,\phi))$ is not needed
to make sense of the stationary phase integration; and therefore
the contribution $Z_\sigma$ of $\widetilde X_\sigma$ to the integral
has a limit as $\epsilon\to 0$.  In fact, general principles of stationary
phase integration assert that $Z_\sigma$ depends only on the behavior
of the integrand in \murmor\ up to finite order near the critical locus
$\widetilde X_\sigma$, and so when $\widetilde X_\sigma$ is supported at
$\phi^a=0$, $Z_\sigma$ is a polynomial in $\epsilon$.

In case $G$ acts freely on $X_\sigma$, one can make this much more explicit.
Under this condition,
$H_G^*(X_\sigma)$ is naturally isomorphic to $H^*(X_\sigma/G)$.
Let ${Y}$ be an equivariant tubular neighborhood of $X_\sigma$
in $X\times {\cal G}$.  ${Y}$ is equivariantly contractible to
$X_\sigma$, so $H_G^*({Y})\cong H_G^*(X_\sigma)\cong H^*(X_\sigma/G)$.
If $\pi:{Y}\to X_\sigma/G$ is the composite of an equivariant
retraction ${Y}\to X_\sigma$ and the natural projection
$X_\sigma\to X_\sigma/G$, then the natural isomorphism of $H^*(X_\sigma/G)$
with $H_G^*({Y})$ is simply the pullback $\pi^*:H^*(X_\sigma/G)\to
H_G^*({Y})$.  In particular, the element
$-(\phi,\phi)/2\in H_G^4(X)$, when restricted to ${Y}$, is
$\pi^*(\Theta)$ for some $\Theta\in H^4(X_\sigma/G)$.
(In fact, $\Theta$ is a characteristic
class of $\mu^{-1}(0)$, regarded as a principal $G$ bundle over
$\mu^{-1}(0)/G$.)
At the level of equivariant forms,
$$-{(\phi,\phi)\over 2}=\pi^*(\Theta)+Dw, \eqn\ffon$$
for some $w\in\Omega_G^3({Y})$.

The fact that $X_\sigma=\widetilde X_\sigma$ means that the contribution
of $X_\sigma$ to $\oint \alpha$ can be evaluated by stationary phase
evaluation of
the integral
$${1\over {\rm vol}(G)}\int {\d\phi_1\d\phi_2\dots \d\phi_s
\over (2\pi)^s}
\alpha\cdot \exp\left({t\,\,\d\lambda-it\sum_a\phi^a\lambda(V_a)}
-{\epsilon\over 2}\sum_a(\phi^a)^2\right)\cdot u, \eqn\lkkl$$
where $u$ is a smooth $G$-invariant
function that is 1 in an equivariant neighborhood
of $X_\sigma$ and zero outside ${Y}$.
In using \ffon, one can integrate by parts and discard the $\d u$ term,
since there are no critical points of $K$ where $\d u\not= 0$.
So in evaluating the contribution $Z_\sigma$ of $X_\sigma$,
we can make the substitution
$$ \exp\left(
-{\epsilon\over 2}(\phi,\phi)\right)\to \exp({\epsilon}\Theta)
  .\eqn\bobo$$
Among other things, this makes it clear that $Z_\sigma$ is a polynomial
in $\epsilon$ of order at most ${1\over 4}{\rm dim}(X_\sigma/G)$.

Moreover, the isomorphism $H_G^*({Y})\cong H^*(X_\sigma/G)$ means
that any $\alpha\in H^*_G(X)$ that we may be trying to integrate,
when restricted to a neighborhood of $X_\sigma$, is the pullback
of some $\alpha'\in H^*(X_\sigma/G)$.  So we can make the substitution
$$\alpha\to \alpha' \eqn\uxux$$
in the integral.  We will later find an important situation to which
these considerations apply.

\subsection{Derivation Of The DH Formula}

For clarity,
we will now recall [\bismut] how to obtain the DH formula
in a similar fashion.
So we assume $G=U(1)$.  The differential in the de Rham model
of $H_G^*(X)$ is now $D=\d-i\phi j_V$, where $V$ is a vector field
generating the $G$ action, and $\phi$ is a linear function on the
one dimensional Lie algebra of $G$.  We ``localize'' algebraically
by setting $\phi = i$.  An equivariant form is then simply a differential
form $\alpha$ obeying $(\d+j_V)\alpha=0$.  By integration of such
a form we simply mean integration over $X$ in the usual sense.

We have
$$\int_X \alpha =\int_X \alpha\cdot \exp(tD\lambda), \eqn\hddx$$
for any $G$-invariant $\lambda$.  To pick a suitable $\lambda$,
let $g$ be a $G$-invariant Riemannian metric on $X$, and let $\lambda$
be the one-form $\lambda=-g(V,\cdot)$.  Then
\hddx\ amounts to
$$\int_X\alpha =\int_X\alpha\cdot \exp(-t\,\,\d\lambda-tg(V,V)).\eqn\ddx$$
So taking $t\to \infty$, we get a localization at the zeros of $g(V,V)$
or in other words at the zeros of $V$.  At an isolated zero $P$ of $V$,
the Hessian of $g(V,V)$ is non-degenerate, so the large $t$ limit
can be evaluated by Gaussian integration.  In this way, one gets the
contribution of $P$ to the generalized DH formula of [\abot, eqn.
(3.8)].

\section{The Symplectic Case}

We now want to elucidate the meaning of the localization formula \mirrp\
in the following important case.  We suppose that $X$ is a symplectic
manifold, with symplectic form $\omega$, and that the $G$ action on $X$
has a moment map $\mu$.  We pick on $X$ an almost complex structure
$J$ such that $\omega$ is of type $(1,1)$ and positive.
Positivity means that the metric $g(\cdot,\cdot)$ defined by
$$g(u,v)=\omega(u,Jv) \eqn\onbo$$
is positive definite.  Such a $J$ always exists (and is unique up to homotopy)
because the Siegel upper half plane is contractible.
Set $I=(\mu,\mu)$ and
$$\lambda={1\over 2}J(dI).      \eqn\turno$$
(So in components, $\lambda={1\over 2}J^i{}_j\partial_iIdx^j$.)

Obviously, at the critical points of $I$, $\lambda=0$ and
hence $\lambda(V_a)=0$.  We want to prove the converse.
Let $Y=\sum_a\mu_aV_a$.  The moment equation $\d\mu_a=-j_{V_a}(\omega)$
implies that $Y={1\over 2}\omega^{-1}dI$, where $\omega^{-1}$
(which in components
is the inverse matrix to $\omega$, so $(\omega^{-1})^{ik}\omega_{kj}=
\delta^i_j$) is regarded as a map $T^*X\to TX$.
$\lambda(V_a)=0$ implies $\lambda(Y)=0$ or
$$\omega^{-1}(\d I,J\d I)=0.     \eqn\gurko$$
Positivity of \onbo\ means that \gurko\ holds only for $\d I=0$.
Thus, in the case of symplectic manifolds, with our choice of $\lambda$,
the localization principle
is a reduction to the critical points of $I=(\mu,\mu)$.

\subsection{$G$ Action On $T^*G$}

We now consider the basic case of
$X=T^*G$, with the natural symplectic structure on $T^*G$ and the
$G$ action on $T^*G$ induced from the right action on $G$.
The localization will be on $G\subset T^*G$.
As $T^*G$ is not compact, we must define the $\oint$ operation by the right
hand side of \urmor, insisting on $t\not= 0$.
It is really the local behavior near $G\subset T^*G$
that matters, and the importance of the calculation is that it can be
applied whenever the critical point set of $I$ has a component that can be
modeled on $G\subset T^*G$.

By going to a basis of right-invariant one forms, $T^*G$ can be identified
with $G\times{\cal G}$.
In terms of $g\times \gamma\in G\times {\cal G}$,
the symplectic form can be written
$$\eqalign{\omega &=(\d\gamma,\d gg^{-1})+(\gamma,\d gg^{-1}\d gg^{-1})
\cr &= (\d\gamma+{1\over 2}[\gamma,\d gg^{-1}],\d gg^{-1}). \cr}\eqn\ucco$$
The vector field $V$ associated with $a\in{\cal G}$ can be described
by the formulas
$$\delta g=-ga,~~~~~~~~~~~\delta\gamma=0. \eqn\uccx$$
A small computation gives $j_V(\omega)=\d(\gamma,gag^{-1})$, so the moment
map is $\mu(a)=-
(\gamma,gag^{-1})$.  The square of the moment map $I=(\mu,\mu)$
is therefore $I=(\gamma,\gamma)$.
An almost complex structure on $T^*G$ can be defined by the formulas
$J(\theta)=\eta$, $J(\eta)=-\theta$, with
$$\eqalign{\theta & =\d\gamma+{1\over 2}[\gamma,\d gg^{-1}] \cr
\eta & = \d g g^{-1}.\cr}
\eqn\gogo$$
It is evident from the second description of $\omega$ in \ucco\ that
$\omega$ is positive and of type $(1,1)$.
One now computes that $\lambda={1\over 2}J(\d I)$ is
$$\lambda=(\gamma,\d gg^{-1}).        \eqn\ccx$$

$G$ acts freely on $T^*G$; the quotient can be identified with ${\cal G}$
via the projection $G\times {\cal G}\to{\cal G}$.  So the equivariant
cohomology of $T^*G$ coincides with the ordinary cohomology of ${\cal G}$,
and vanishes except in dimension zero, where it is represented by the
constants.  The only integral that we really have to consider is therefore
$\oint_{T^*G} 1$.  This is
$$\oint_{T^*G} 1={1\over {\rm vol}(G)}\int {\d\phi_1\dots \d\phi_s\over
(2\pi)^s}\int_{T^*G} \exp\left(
{tD\lambda}-{\epsilon\over 2}(\phi,\phi)\right).
\eqn\uuurgo$$
Since the integral is independent of $t$ for $t\not=0$, we can set $t=1$.
Working out the top form component of $D\lambda$ explicitly,
and changing variables from $\phi$ to $g^{-1}\phi g$, we get
$$\oint_{T^*G} 1 ={1\over {\rm vol}(G)}\int {\d\phi_1\dots \d\phi_s\over
(2\pi)^s}\int_{T^*G} \exp\left({-i(\gamma,\phi)-{\epsilon\over 2}(\phi,\phi)}
\right)  {(\d\gamma,\d gg^{-1})^n\over n!}.\eqn\hipd$$
(Note that $\omega^n$ coincides with $(\d\gamma,\d gg^{-1})^n$.)
The $\gamma$ integral can be done using
$$\int_{-\infty}^\infty \d u\,\,e^{-iuv}=2\pi\delta(v).\eqn\ipd$$
The $\phi$ integral can then be done using the delta functions.  Notice
that the $\epsilon$ dependence disappears at this point
(a consequence of \bobo).  The integral
over $g$ cancels the factor of ${\rm vol}(G)$.  Assuming that the orientation
of $T^*G$ is chosen to agree with the one determined by the symplectic
structure, the result is just
$$\oint_{T^*G} 1 = 1. \eqn\ussu$$

That this is independent of $\epsilon$ is a consequence of
the fact that for $X=T^*G$, \mcmc\ implies $\phi=0$
(this is shown more generally in (2.45) below).
Hence we could alternatively have proceeded as follows:
set $\epsilon$ to zero in \hipd; evaluate the $\phi$ integral using
\ipd; evaluate the $\gamma$ integral using the resulting delta functions.

\subsection{$G$ action on $T^*(G/H)$}

We will now briefly
describe the extension of this to the case that $X=T^*(G/H)$,
with $H$  a subgroup of $G$ of dimension $r$ and Lie algebra ${\cal H}\subset
{\cal G}$, and with the
natural symplectic structure on $T^*(G/H)$.\foot{This case is generally
less important
in our applications, but will arise when we consider gauge theory in genus
zero.}
The restriction of
$(~,~)$ to ${\cal H}$ is an invariant quadratic form on ${\cal H}$
that we will denote by the same symbol.
We have already carried out equivariant integration over $T^*(G/H)$ for
$H=G$ and $H=\{1\}$ in \pointtarget\ and
\ussu\ respectively.  The general case turns out to be a natural
combination of these.

$T^*(G/H)$ can be parametrized by pairs $g\times\gamma\in G\times {\cal G}$,
with an equivalence relation
$g\times \gamma \cong hg\times h\gamma h^{-1}~{\rm  for}~h\in H,$
and a constraint
$(\gamma,b)=0$, for $b\in {\cal H}$.   The symplectic structure
and the $G$ action can still be described by \ucco\ and \uccx.
The moment map for $a\in {\cal G}$
is still $\mu=-(\gamma,gag^{-1})$, and its square is still
$I=(\mu,\mu)=(\gamma,
\gamma)$.

Let $b_i$ be an orthonormal basis of ${\cal H}$,
let $\Pi$ be the orthogonal projection onto the
complement ${\cal H}_\perp$ of ${\cal H}$,
and introduce the ${\cal H}_\perp$-valued one-forms
$$\eqalign{\theta & = \Pi\left(\d\gamma+{1\over 2}[\gamma,\d gg^{-1}]
     +{1\over 2}\sum_i[\gamma,b_i](b_i,\d gg^{-1})\right) \cr
           \eta   & = \Pi \left(\d gg^{-1}\right) . \cr} \eqn\jsjsj$$
An almost complex structure $J$ such that $\omega$ is positive and of type
$(1,1)$ can be defined as before
by the formulas  $J(\theta)=\eta,\,\,\,J(\eta)=-\theta$.

Now we introduce the usual $G$-invariant one-form
$$\lambda = {1\over 2} J\d\mu^2={1\over 2}J\d(\gamma,\gamma)=(\gamma,\d
gg^{-1}).
      \eqn\xonxno$$
So
$$ D\lambda = (\d\gamma,\d
gg^{-1})+(\gamma,\d gg^{-1}\d gg^{-1})+i(\gamma,g\phi g^{-1})
      \eqn\udu$$

Let $\alpha$ be an arbitrary element of $H^*_G(T^*(G/H))$.  Since
$T^*(G/H)$ has an equivariant retraction to $G/H$, we can represent
$\alpha$ by the pullback of an element of $\Omega^*_G(G/H)$, which
we will also call $\alpha$.  We want to compute
$$\oint_{T^*(G/H)}\alpha  ={1\over {\rm vol}(G)}\int{\d\phi_1\dots
\d\phi_s\over (2\pi)^s}\int_{T^*(G/H)}\alpha\cdot \exp\left(
tD\lambda-{\epsilon\over 2}(\phi,\phi)
\right).  \eqn\xfxnnn$$
Using \udu, setting $t$ to 1,
extracting the top form component, and replacing $\phi$ by
$g^{-1}\phi g$, this is
$$\oint_{T^*(G/H)}\alpha = {1\over {\rm vol}(G)}\int {\d\phi_1\dots
\d\phi_s\over
(2\pi)^s}\cdot {(\d\gamma,\d gg^{-1})^s\over s!}\alpha
\exp\left(i(\gamma,\phi)-{\epsilon\over 2}
(\phi,\phi)\right). \eqn\mcmcmmc$$
$\alpha$, being a pullback from $G/H$, is independent of $\gamma$.
If, therefore, we write $\phi=\phi'+\phi_\perp$, with $\phi'\in {\cal H}$ and
$\phi_\perp\in {\cal H}_\perp$, then the $\gamma$ integral
gives $(2\pi)^r\delta^r(\phi_\perp)$, with the aid of which the $\phi_\perp$
integral can be done.  It remains to integrate over $\phi'$ and $g$.

Restricted to a given point in $G/H$, say the coset of $1\in G$, $\alpha$
reduces to an $H$-invariant polynomial $\alpha'(\phi')$.
(The map $\alpha\leftrightarrow\alpha'$ is a bijection in cohomology and
is the natural isomorphism $H_G^*(G/H)\cong H_H^*({\rm pt})$.)
The $\phi'$
integral gives a factor
$$\int {\d\phi'_1\dots \d\phi'_{s-r}\over (2\pi)^{s-r}}\cdot \alpha'(\phi')
\cdot
\exp(-{\epsilon\over 2}(\phi',\phi')). \eqn\sis$$
The $g$ integral gives a factor of ${\rm vol}(G/H)
={\rm vol}(G)/{\rm vol}(H)$.  (The volumes are computed using the measures
on $G$, $H$, $G/H$ induced from the quadratic form $(~,~)$ on ${\cal G}$
and ${\cal H}$.)
Combining the pieces, and assuming that $T^*(G/H)$ is given the orientation
compatible with its symplectic structure, we get
$$\oint_{T^*(G/H)}\alpha={1\over {\rm vol}(H)}\int{\d\phi'_1\dots
\d\phi'_{s-r}
\over (2\pi)^{s-r}}\cdot \alpha'(\phi')\cdot
\exp\left(-{\epsilon\over 2}(\phi',\phi')\right)
,\eqn\dnd$$
which shows that $G$-equivariant integration on $T^*(G/H)$ reduces to
$H$-equivariant integration on a point.

\section{More On The Symplectic Case}

We now want to make a more intensive study of equivariant integration
on a symplectic manifold $X$ with symplectic  form $\omega$.
We have seen that this reduces to
a sum over critical points of the function $I=(\mu,\mu)$.
The absolute minimum of that function is $\mu^{-1}(0)$.
Assume first that $\mu^{-1}(0)$ is a smooth manifold, on which $G$
acts freely, so that   the quotient ${\cal M}=\mu^{-1}(0)/G$ is a smooth
manifold with a naturally induced symplectic structure, which we will also
call $\omega$.

We want to compute the contribution of $\mu^{-1}(0)$ to
$$\oint_X\alpha={1\over {\rm vol}(G)}\int{\d\phi_1\dots \d\phi_s\over
(2\pi)^s}\int_X\alpha\cdot \exp\left(tD\lambda
-{\epsilon\over 2}(\phi,\phi)\right).
\eqn\snsnss$$
This reason that this is simple is that we can use \bobo\ and \uxux.
We recall that the criteria for the validity of \bobo\ were that $G$
should act freely on the critical point set, and that
$$\sum_a\phi^a\,\,\d\left(\lambda(V_a)\right) = 0 \eqn\mcmcmcmc$$
should imply $\phi^a=0$.  A small computation shows that
$\lambda(V_a)=\sum_bg(V_a,V_b)\cdot\mu_b$ where $b$ runs over an orthonormal
basis of ${\cal G}$, and $g(\cdot,\cdot)$ is
the metric \onbo, which is positive definite by the choice of $J$.
Wherever $G$ acts freely, the $V_a$ are linearly independent, and $g(V_a,V_b)$
is therefore a positive definite metric on ${\cal G}$ (but not
necessarily $G$-invariant,
of course, as we are not at a fixed point of $G$).
At $\mu=0$, \mcmcmcmc\ reduces to
$$\sum_a\phi^a\cdot \sum_b g(V_a,V_b)\cdot \d\mu_b =0.       \eqn\hshss$$
Since $g(V_a,V_b)$ is invertible, and the $\d\mu_b$ are linearly independent
wherever $G$ acts freely (since $\d\mu_b=-j_{V_b}\omega$),
the coefficients of the $\phi^a$ in \hshss\ are linearly independent
on $\mu^{-1}(0)$, and hence also in a neighborhood thereof.
So, as desired, \mcmcmcmc\ implies $\phi^a=0$, and we can use
\bobo\ and \uxux.

\REF\sternberg{V. Guillemin and S. Sternberg, {\it Symplectic Techniques
In Physics,}  (Cambridge University Press, 1984).}
Thus, let $Y$ be an equivariant tubular neighborhood of $\mu^{-1}(0)$.
Pick an equivariant retraction $Y\to \mu^{-1}(0)$.  Composing this
with the natural projection $\psi:\mu^{-1}(0)\to \mu^{-1}(0)/G$, one
gets an equivariant projection $\pi:Y\to \mu^{-1}(0)/G$.
The elements $-(\phi,\phi)/2$ and $\alpha$ of $H^*_G(Y)$
are the pullbacks by $\pi$
of some classes $\Theta$, $\alpha'$ in $H^*(\mu^{-1}(0)/G)$.
\bobo\ and \uxux\ mean that in evaluating
the contribution $Z(\mu^{-1}(0))$
of $\mu^{-1}(0)$ to $\oint_X\alpha$, we can write
$$Z(\mu^{-1}(0))={1\over {\rm vol}(G)}\int{\d\phi_1\dots \d\phi_s\over
(2\pi)^s}\int_Y\alpha'\cdot \exp\left(tD\lambda
+{\epsilon}\Theta\right).   \eqn\hshshz$$
To study this integral, first integrate over the fibers
of $\pi$.  Everything in \hshshz\ is a pullback via $\pi$ except
$\exp(tD\lambda)$.  Hence we must evaluate
$$ \int{\d\phi_1\dots \d\phi_s\over (2\pi)^s}
\int_{\pi^{-1}({\rm pt})} ~\exp
\left(tD\lambda\right).
\eqn\gogg$$
Now, $\pi^{-1}({\rm pt})$ is fibered over $G\cong \psi^{-1}({\rm pt})$,
and the $G$ action on $\pi^{-1}({\rm pt})$ can be modeled on a neighborhood
of $G\subset T^*G$ (if $Y$ and $\pi$ are constructed as explained in
[\sternberg, Theorem 39.2 and Proposition 40.1]).
Hence, the large $t$ limit of \gogg\ is 1, using \ussu.
For instance, one can do the calculation as in the comment after \ussu,
integrating first over $\phi$ in \gogg\ to produce a delta function supported
on $\mu^{-1}(0)$.
So integrating over the fibers of $\mu^{-1}(0)\to \mu^{-1}(0)/G$,
we get the very simple result
$$Z(\mu^{-1}(0))=\int_{\mu^{-1}(0)/G}\alpha'\cdot\exp({\epsilon}
\Theta).        \eqn\pxpxx$$
This formula is a major ingredient in our applications.

\subsection{Higher Critical Points}

We also want to say something about the contributions of the higher
critical points of $I$ in equivariant integration over $X$.
At this point, we must specify what sort of equivariant forms
we wish to integrate.
Whenever one has a Hamiltonian
group action on a symplectic manifold, a basic equivariant differential
form, exploited in [\abot],
is the form of degree two $\overline\omega=\omega-i\sum_a\phi^a\mu_a$.
It can be regarded as an equivariant extension of $\omega$.
For $\alpha$ we will take
$$\alpha=\exp\left(\overline \omega\right)\cdot\beta, \eqn\goodch$$
where we will require that $\beta$ has only a polynomial dependence on $\phi$.
This means that the integrals we wish to calculate are of the general type
(setting $\beta=1$ for simplicity)
$$\oint \alpha={1\over {\rm vol}(G)}\int{\d\phi_1\dots \d\phi_s\over {2\pi}^s}
\exp\left(\omega-i\sum_a\phi^a\mu_a-{\epsilon\over 2}(\phi,\phi)\right)
\eqn\snnsnsnn$$
After performing the $\phi$ integral, this is
$$\oint\alpha
={1\over {\rm vol}(G)\cdot (2\pi\epsilon)^{s/2}}
\cdot\int_X{\omega^n\over n!}\exp\left(-{I\over 2\epsilon}
\right),\eqn\dhdh$$
with $I=(\mu,\mu)$.
We note that, apart from some elementary factors, this is the integral
\juju\ discussed in the introduction.

Let $X_\sigma$ be a component of the critical set, and $Y$ an equivariant
tubular neighborhood of $X_\sigma$.  We want to estimate, for small $\epsilon$,
the contribution $Z(X_\sigma)$ of $X_\sigma$
to \dhdh\ (and its generalization with $\beta\not= 1$).
In doing so we will assume that $X_\sigma$ is a nondegenerate critical
locus in the
extended sense of Bott -- that it is a smooth manifold and
that the Hessian of $I$ is invertible in
the directions normal to $X_\sigma$.
Naively, one would expect that, if there is any sort of representation
of \dhdh\ as a sum over critical points, the contribution of $X_\sigma$
for small $\epsilon$ should be of order
$$Z(X_\sigma)\sim \exp\left(-{I(X_\sigma)\over 2\epsilon}\right),
\eqn\asympbe$$
(where $I(X_\sigma)$ is the constant value of $I$ on $X_\sigma$)
up to a power of $\epsilon$.
This is so; we will see how this behavior emerges from the
general localization procedure.
To begin with, we consider the case $\beta=1$.

$Z(X_\sigma)$ is the large $t$ limit of
$$Z(X_\sigma;t)={1\over {\rm vol}(G)}\int{\d\phi_1\dots \d\phi_s\over (2\pi
)^s}\int_Y \exp\left(\omega+t\,\,\d\lambda -i\sum_a\phi_a(\mu_a
+t\lambda(V_a))-{\epsilon\over 2}(\phi,\phi)\right).\eqn\murok$$
Upon performing the Gaussian integral over $\phi$, this becomes
$${1\over (2\pi\epsilon )^{s\over 2}
}\int_Y  \exp\left(\omega+t\,\,\d\lambda-{1\over 2\epsilon}W\right).
\eqn\gufus$$
where
$$ W=\sum_a(\mu_a+t\lambda(V_a))^2=\sum_a(\mu_a+t\sum_bg(V_a,V_b)\mu_b)^2
.\eqn\jdoxx$$

Restricted to the critical component $X_\sigma$ of $I=(\mu,\mu)$,
$W$ is equal to the constant $I(X_\sigma)$,
since $\sum_bg(V_a,V_b)\mu_b=0$
on critical points.
We will show momentarily that for $t>>0$,
$X_\sigma$ is a local minimum of $W$ (even though it may be an unstable
critical set of $I$), and that
the Hessian in the normal directions is positive
definite and of order $t$.
The integral in \gufus\ is sharply peaked around this
minimum, so the large $t$ behavior
of \gufus\ is determined by local behavior near
the $X_\sigma$.\foot{But a Gaussian approximation
to the integral near $X_\sigma$
is generally not valid, because the $t^2W''$ term vanishes up to fourth order
yet -- because of the $t^2$ factor -- cannot be ignored.
For instance, the example treated in the appendix
gives a two dimensional integral that is roughly $\int \d x_1\,\,\d x_2
\exp\left(-t(x_1^2+x_2^2)-t^2(x_1^2+x_2^2)^2\right)$, where
$x_i$ are local coordinates centered at one of the critical points.
This integral
cannot be approximated for large $t$ by a Gaussian, and instead
is easily seen to give the error function found in the appendix.
The terms proportional
to $t$ and $t^2$ correspond to $W'$ and $W''$.}
The large $t$ limit of \gufus\ therefore vanishes
exponentially for $\epsilon\to 0$ as $\exp\left(-W(X_\sigma)/2\epsilon
\right)=\exp\left(-I(X_\sigma)/2\epsilon\right)$.
This is the exponential behavior suggested intuitively in \asympbe.
The same exponential would arise if we consider
not $\oint \exp(\overline \omega)$, but
$\oint\exp(\overline\omega)\cdot \beta$, with $\beta$ an arbitrary
equivariant form with a polynomial dependence on $\phi$.  The introduction
of $\beta$ would not modify the exponential factor in \gufus; it would
merely produce a prefactor behaving for small $\epsilon$ as $\epsilon^{-n}$
for some $n$.

It remains to show that for large enough $t$, $X_\sigma$ is a local
minimum of $W$, with a Hessian of order $t$.
Let $\widehat V=\sum_aV_a\mu_a=\omega^{-1}\d I/2$.
We can write
$$W = I+W'+W'' \eqn\rifo$$
with
$$\eqalign{W'&=2tg(\widehat V,\widehat V) \cr W''& =
t^2\sum_a(g(V_a,\widehat V))^2.\cr}\eqn\snsnzz$$
At a critical point of $I$, $\widehat V=0$, so such a point is also
a critical point of $W'$ and $W''$.
$W'$ and $W''$ are positive semidefinite and vanish precisely at critical
points of $I$.

At a critical point of $I$, the Hessian or matrix of second derivatives
$\partial^2I/\partial x^i\partial x^j$ may not be positive definite.
However, it is dominated
for $t\to +\infty$ by the Hessian of $W'$, which turns out to be
$${\partial^2W'\over\partial x^i\partial x^j}={t\over 2}{\partial^2I
\over\partial x^i\partial x^k}g^{kl}{\partial^2I\over\partial x^l\partial x^j}
\eqn\fgghd$$
and in particular is positive semidefinite
(here $g^{kl}$ are the matrix elements of the inverse of the metric
$g(\cdot,\cdot)$).
As critical points of $I$ are certainly local minima of $W''$,
this suffices to show that the Hessian of $W$ is positive semidefinite
(and proportional to $t$ for large $t$)
and that its kernel consists at most of the kernel of the Hessian of $I$.
Since we have assumed that $X_\sigma$ is a nondegenerate critical
locus of $I$ (in the sense that the Hessian is invertible in the normal
directions), $X_\sigma$ is a local minimum
of $W$.

\chapter{Rederivation In Quantum Field Theory Language}

\REF\wwitten{E. Witten, ``Topological Quantum Field Theory,''
Commun. Math. Phys. {\bf 17} (1988) 353,
``Introduction To Cohomological Field Theory,''
Int. J. Mod. Phys. {\bf A6} (1991) 2775.}
\REF\kanno{H. Kanno, ``Weil Algebra Structure And Geometrical Meaning Of The
BRST Transformation In Topological Quantum Field Theory,'' Zeitschrift
fur Physik C - Particles And Fields {\bf 43} (1989) 477.}
\REF\bais{P. van Baal, ``An Introduction To Topological Yang-Mills Theory,''
Acta. Phys. Polon. {\bf B21} (1990) 73.}
\REF\atj{M. F. Atiyah and L. Jeffrey, ``Topological Lagrangians And
Cohomology,'' J. Geom. Phys. {\bf 7} (1990) 119.}
\REF\baulieu{L. Baulieu and I. M. Singer, ``Topological Yang-Mills
Symmetry,'' Nucl. Phys. (Proc. Supp.) {\bf 5B} (1988) 12.}
\REF\birm{D. Birmingham, M. Blau, M. Rakowski, and G. Thompson,
``Topological Field Theory,'' Phys. Reports {\bf 209} (1991) 129.}
\REF\uvwitten{E. Witten, ``The $N$ Matrix Model And Gauged WZW Models,''
IAS preprint, to appear in Nucl. Phys. B.}
In this section, I will rederive some of
the main results of the last section,
using this time the language of quantum field theory.
To begin with, I want to describe the nonabelian localization
principle using more physical language, running
through the relevant portions of \S2 more quickly with the notation
of physicists.

Notation is generally as in \S2.  In particular,
$X$ is once again a compact
manifold acted on by a compact Lie group $G$, with Lie algebra ${\cal G}$;
the $G$ action on $M$ is generated by vector fields $V_a$.

We let $x^i$ be local coordinates on $M$, and let $\psi^i$ be anticommuting
variables tangent to $M$.  Let $\phi^a$ be bosonic variables in the
adjoint representation of $G$.  Equivariant differential forms
(elements of $\Omega_G^*(X)$) are just
$G$-invariant functions of $x,\psi,\phi$.  Let
$$D=\sum_i\psi^i{\partial\over \partial x^i}-i\sum_{i,a}\phi^a
V_a^i{\partial\over\partial\psi^i}.          \eqn\nnsn$$
This operator
obeys $D^2=0$ when acting on $G$-invariant functions of $x,\psi,\phi$.
$D$ is the standard differential in $\Omega_G^*(X)$, written
in physical notation.
Integration of an equivariant differential form $\alpha$ is defined by
$$\oint \alpha={1\over {\rm vol}(G)}\int_{X}\d x^i\d\psi^i
\int_{\cal G}{\d\phi_1\dots \d\phi_s\over (2\pi)^s} \alpha \cdot \exp
\left(-{\epsilon\over 2}
(\phi,\phi )\right). \eqn\nnnsx$$
Note that although there is no natural measure for the $x$'s or $\psi$'s
separately, there is a natural measure $\d x^i\d\psi^i$, since the Jacobian
in a change of variables on $M$ would cancel between bosons and fermions.
(Integration with respect to this measure is what mathematicians call
integration of differential forms.)
As in \S2, $\d\phi_1\dots \d\phi_s$ is an arbitrary measure on ${\cal G}$,
and (identifying the tangent spaces of $G$ with ${\cal G}$) the same
measure is used in computing ${\rm vol}(G)$.
Obviously, by integrating by parts in $x$ and $\psi$,
$$\oint D\beta =0, \eqn\snnns$$
for any $\beta$.

Nonabelian localization comes from the fact that -- if $D\alpha=0$ --
$$\oint \alpha =\oint \alpha \exp(tD\lambda), \eqn\nnns$$
for any equivariant form $\lambda$.  This follows from \snnns, using
$D\alpha= D^2=0$.
We pick
$$\lambda=\sum_i\psi^ib_i, \eqn\xxcx$$
where $b_i$ is a function of the $x$'s only.  (So in mathematical
terminology, $\lambda=
\sum_ib_i\d x^i$ is the $G$-invariant one-form on $M$ used in \S2.)
We insert this in \nnnsx, compute $D\lambda$,
and perform the $\phi$ integral.  If $\alpha$ is independent of
$\phi$, we get
$$\oint\alpha = {1\over {\rm vol}(G)\cdot (2\pi\epsilon)^{s/2}}
\int_X\d x^i\d\psi^i \,\,\,\alpha \exp\left(t\sum_{i,j}
\psi^i\psi^j\partial_ib_j\right)
\cdot\exp\left(-{t^2\over 2}\sum_a(V_a^ib_i)^2\right). \eqn\xcx$$
If $\alpha$ depends on $\phi$ in a sufficiently mild fashion, then  \xcx\
is replaced by a possibly more complicated formula with similar properties.
The main point of \xcx\ is that in the limit of $t\to\infty$, the integral
becomes localized near the solutions of
$$V_a^ib_i = 0, ~~~~~a=1\dots s, \eqn\turgo$$
and can be written as a sum of contributions that depend only on the local
data near solutions of \turgo.  This is the nonabelian localization of \S2.

Now, in our more detailed applications, we wish to assume that
$X$ is a symplectic manifold, with symplectic form $\omega={1\over
2}\omega_{ij}
\psi^i\psi^j$.  We also assume that the $V_a$ are derived via Poisson brackets
from Hamiltonian functions $\mu_a$, or in other words that
$$V_a^i=\omega^{ij}\partial_j\mu_a        \eqn\uuuurgo$$
(where $\omega^{ij}\omega_{jk}=\delta^i_k$).
In this case, we
set $I=\sum_a\mu_a^2$, and we let
$$b_j={1\over 2}J^i{}_j\partial_iI, \eqn\rgo$$
with  the matrix $J$ restricted by requiring that $J^2=-1$ and
that the ``metric''
$$ g_{ij}=J^k{}_i\omega_{kj} \eqn\jjddx$$
is symmetric and positive definite.  This is usually described by saying
that $J$ is an almost complex structure on $X$
for which $\omega$ is of type $(1,1)$
and positive.  $J$'s obeying these conditions always exist.
In this situation, \turgo\  can be written
$$ 0=V_a^i\partial_i I. \eqn\hurgo$$  This implies that
$$0 = \sum_a\phi^aV_a^ib_i=g^{ij}\partial_iI\partial_j I,       \eqn\rrgo$$
and hence that
$$0=\partial_iI.        \eqn\ddgo$$
Since \ddgo\ obviously implies \hurgo, we have learned that, with the
particular choice of $\lambda$ that we have made, the nonabelian localization
for Hamiltonian actions on symplectic manifolds  amounts to a formula involving
a sum over the critical points of $I$, another main result of \S2.

In \S2, we then specialized further to the case that
$$\alpha=\exp\left({1\over 2}\omega_{ij}\psi^i\psi^j-i\phi^a\mu_a\right)
\cdot \beta, \eqn\ssxxx$$
where $\beta$ has at most polynomial dependence on $\phi$.
For instance, suppose $\beta=1$.
In this case, it is convenient to perform the $\psi$ integral
to reduce the integration over $x,\psi$ to an integration over $x$ only.
In fact, although in general there is no natural measure on a manifold $X$,
by introducing the symplectic form $\omega$ as in \ssxxx\ and performing the
$\psi$
integral we get such a measure:
$$ \d x^1\wedge \dots\wedge \d x^n\int \d\psi^1\dots
\d\psi^n\,\,\,\exp({1\over 2}\omega_{ij}\psi^i\psi^j). \eqn\naturalmeasure$$
As the $\psi$ integral gives $\sqrt {\det \omega}$,
the measure on $X$ obtained by performing this integral
is simply the standard Liouville measure, which
if $\omega$ is regarded as a two-form is usually written $\omega^n/n!$
In any event, up to a constant multiple, this is the only measure on $X$ that
can be constructed using $\omega$ alone.

After performing also the Gaussian integral over $\phi$, we get
$$\oint\alpha={1\over {\rm vol}(G)\cdot (2\pi\epsilon)^{s/2}}
\int_X{\omega^n\over n!}
\exp\left(-{1\over 2\epsilon}\left(\mu,\mu\right)\right).  \eqn\sxxs$$
A more elaborate expression with the same essential properties arises
if we permit $\beta\not=1$ (with polynomial $\phi$ dependence).

\sxxs\ makes it clear that the dominant contribution for $\epsilon\to 0$
comes from the absolute minimum of $I=(\mu,\mu)$, that is the
solutions of $\mu=0$.  One also expects heuristically that, in the evaluation
via nonabelian localization, an arbitrary critical
point $P$ must make a contribution of order $\exp\left(-I(P)/2\epsilon\right)$
for $\epsilon\to 0$.  This latter assertion was justified (under some
assumptions) in \S2.4, and will not be reconsidered here.  What we want
to do here is to reexamine from a physicist's point of view the relation
found in \S2 between the contribution of the minimum at $\mu=0$ and
the  cohomology of ${\cal M}=\mu^{-1}(0)/G$.  This will be done by (in a field
theory language)
mapping a suitable ``cohomological'' field theory (for background see
[\wwitten--\uvwitten])   to a suitable ``physical'' field theory.
The mapping between the two is essentially the proof of nonabelian
localization,
looked at in reverse.  The mapping we will find has an analog, much less
understood, in the relation [\uwitten--\kontsevich]
between physical and topological gravity in
two dimensions.

\section{The Cohomological Gauge Theory}

At this stage, we will specialize to the case of two dimensional gauge
theories.  Thus, for the space $X$ of the above discussion, we take the
space ${\cal A}$ of connections on a vector bundle $E$, with compact
structure group $H$, over
a two dimensional surface $\Sigma$.  For $G$ we take the group of gauge
transformations of $E$.  The gauge field $A$ plays the role of
the $x$'s in the above   formulas; the $\psi$'s are now an anticommuting
one-form with values in the adjoint representation of $H$; and the $\phi$'s
are a zero-form on $\Sigma$
also with values in the adjoint representation.

The $(A,\psi,\phi)$ system is the basic multiplet of cohomological Yang-Mills
theory.  In physical notation the transformation laws are
$$\eqalign{  \delta A_i & = i\epsilon\psi_i       \cr
             \delta \psi_i & =-\epsilon D_i\phi =-
\epsilon \left(\partial_i\phi
               +[A_i,\phi]\right)    \cr
             \delta\phi    & = 0,\cr}    \eqn\hdnn$$
with $\epsilon$ an anticommuting parameter.
In terms of the operator $D$ of equation \nnsn, this can be written
$\delta\Phi=iD\Phi$, for every $\Phi$.  It is also conventionally written
$\delta\Phi=-i\{Q,\Phi\}$, where $Q$ is the BRST operator (so $Q=-D$).
As in the general discussion, $Q^2=0$ (or $D^2=0$) up to a gauge
transformation.
In fact, $Q^2=-i\delta_\phi$, where $\delta_\phi$ is the generator of
a gauge transformation with infinitesimal parameter $\phi$.
We introduce a ghost number quantum number, with the ghost numbers
of $(A,\psi,\phi)$ being $(0,1,2)$.

Additional multiplets, which typically are
needed to write Lagrangians (and which are analogous to antighost
multiplets in usual BRST quantization), can be introduced
in the following standard way.  One considers pairs $(u,v)$ of opposite
statistics and ghost numbers $(n,n+1)$, for some $n$, and with
$$\eqalign{  \delta u & = i\epsilon v \cr
             \delta v & = \epsilon [\phi,u] .\cr}\eqn\msmms$$
In two dimensional gauge theories, it is convenient to introduce
two such pairs $(\lambda,\eta)$ and $(\chi,-iH)$, with $\lambda$ a commuting
field of ghost number $-2$, and $\chi$ an anticommuting field of ghost
number $-1$.  So
$$\eqalign{\delta\lambda & =i\epsilon\eta,~~~~~\delta\eta =\epsilon[\phi,
         \lambda]          \cr
       \delta\chi & =\epsilon H ~~~~~\delta H =i\epsilon[\phi,\chi].\cr}
     \eqn\undox$$

Any expression $L=-i\{Q,V\}$, with $V$ gauge invariant, will be $Q$-invariant,
since $Q^2=0$ on gauge invariant functions.
In writing a Lagrangian, we wish also to pick $V$ so that all fields will have
a nondegenerate kinetic energy.  A suitable choice of $V$, which conserves
ghost number, is
$$V={1\over h^2}\int_\Sigma \d\mu\Tr\left({1\over 2}\chi(H-2\star F)+g^{ij}
D_i\lambda\psi_j\right). \eqn\nsnns$$
Here $\Sigma$ is the Riemann surface on which we formulate the theory.
It has been endowed with a metric $g$, and $\mu$ is the corresponding
Riemannian measure.
$h$ is a real constant.
$F=\d A+A\wedge A$ is the Yang-Mills
field strength, and $\star$ is the Hodge star operator; we also set $f=\star
F={1\over 2}\epsilon^{ij}F_{ij}$.
One finds
that
$$\eqalign{L=-i\{Q,V\}={1\over h^2}&\int_\Sigma \d\mu\Tr\left({1\over 2}
(H-f)^2 -{1\over 2}f^2 -i\chi \star D\psi +iD_i\eta\psi^i
\right.\cr &~~~~~~~~~~~~
  \left.
+D_i\lambda D^i\phi+{i\over 2}\chi[\chi,\phi]+i[\psi_i,\lambda]\psi^i\right)
.\cr}
\eqn\mdmdmmd$$
One can solve for the auxiliary field $H$ by its equation of motion
$$ H =f, \eqn\snnsnsnnn$$
and delete the $(H-f)^2$ term from the Lagrangian.

The quantum field theory with Lagrangian \mdmdmmd\ is a topological field
theory, and independent of the choice of the coupling parameter
$h$, because although the coupling and the metric $g$ appear in $L$,
they only appear in terms of the form $\{Q,\dots\}$ (since $L$ itself
is of this form).  By dropping the $(H-f)^2$ term and taking $h\to 0$,
one sees that all computations can be performed by expanding around the
minimum
$$ F= 0\eqn\sjsnns$$
of the gauge boson kinetic energy.

In the weak coupling limit, the $\phi - \lambda$ integral can be treated
formally via one loop determinants and Feynman diagrams.  If one wishes
to treat this integral more honestly, one either considers $\phi$ to be
complex and $\lambda=\overline\phi$, or (as was natural in [\atj]) one
takes $\phi$ real and $\lambda$ imaginary.  For our purposes, such
a choice need not be specified.
To minimize the scalar kinetic energy
 one requires
$$ 0 = D_i\phi  \eqn\pps$$
(if $\lambda=\overline \phi$; in the other case this requirement comes
from stationary phase).
If $A$ is an irreducible solution of \sjsnns\ (in the sense that its holonomy
group commutes only with the center of $H$), then
\pps\ implies $\phi=0$, since otherwise the holonomy group would have to
commute
with $\phi$.

More fundamentally, \sjsnns\ and \pps\ should be regarded as the conditions
$\delta\chi=0$ and $\delta \psi=0$ for a BRST fixed point, as explained
in [\uvwitten, \S3.1].  ($\delta\chi=0$ coincides with \sjsnns\ after using
\snnsnsnnn.)  Let ${\cal U}$ be the space of solutions of \sjsnns\ and \pps.

\mdmdmmd\ is a ``standard'' Lagrangian for the two dimensional
analog of Donaldson theory.
Because of the independence of $h$, all calculations can be performed
in the weak coupling limit, where as just indicated
they reduce to integrals over ${\cal U}$.  For gauge groups $H$ and
bundles $E$ such that
reducible solutions of \sjsnns\ do not exist, ${\cal U}$ is the same
as the moduli space ${\cal M}$ of flat connections
on $E$ up to gauge transformation, and the correlation functions
are intersection pairings on ${\cal M}$, as analyzed in detail
by Baulieu and Singer [\baulieu] (in the analogous four dimensional theory).
The principal difficulty
in understanding the theory, as explained in the second paper in [\wwitten],
comes from the zero modes of $\phi$ and of $\lambda=\overline \phi$
that can
arise
for reducible connections, and as a result of which ${\cal U}$ and ${\cal M}$
do not coincide in general.  This motivated the
search for the following method of eliminating these fields from the problem.

\section{Mapping To The Physical Theory}
\subsection{The Strategy}

If we replace $V$ by $V+tV'$, with $t$ a constant and $V'$ some new gauge
invariant operator, then the theory with Lagrangian
$$L(t)=-i\{Q,V+tV'\} \eqn\murgo$$
is independent of $t$ as long as (i) $V'$ is such that $L(t)$ has
a nondegenerate kinetic energy for all $t$\foot{This is a quantum field
theory analog of requiring that a family of operators be elliptic.};
(ii) the perturbation by $V'$ does not permit any new fixed points,
that is solutions of $\delta\chi=\delta\psi=0$, to flow in from infinity.
The latter condition is needed because the space of fields over
which one integrates (in performing the Feynman path integral) is not
compact, and all statements about topological invariance require pinning
down the behavior at infinity.

We will actually consider a choice of $V'$ such that condition (ii) is not
obeyed.  When this is so, the fixed point equation $\delta\chi=\delta\psi=0$,
in addition to having the component ${\cal M}$ (or in general ${\cal U}$)
discussed above\foot{Which in
general might
undergo some perturbations when $t$ is varied; but we will arrange to avoid
this.}
will acquire additional components ${\cal M}_\alpha$.  The Feynman path
integral will thus reduce to a {\it sum} over the contributions of ${\cal M}$
and ${\cal M}_\alpha$.
The standard BRST arguments show that the contribution of ${\cal M}$ is
independent of $t$, but as the path integral will give naturally a sum of
the contributions from ${\cal M}$ and ${\cal M}_\alpha$,
it might appear that there is no way from studying the (simpler)
$L(t\not= 0)$ theory to recover the result of the (interesting but harder)
theory $L(t=0)$.  To accomplish this, we will at a judicious moment
introduce one further
trick to disentangle the contributions of ${\cal M}$ and ${\cal M}_\alpha$.

\subsection{Elimination of $\lambda$}

We set
$$V'=-{1\over h^2}\int_\Sigma \d\mu\Tr \chi\lambda. \eqn\dogo$$
This will lead to a Lagrangian that does not conserve ghost number;
the ghost number of $V'$ is $-3$, so that of $\{Q,V'\}$ will be $-2$.
We find that the analog of \mdmdmmd\ is
$$\eqalign{L(t)=&-i\{Q,V+tV'\}={1\over h^2}\int_\Sigma \d\mu\Tr\left({1\over 2}
(H-\lambda t-f)^2 -{1\over 2}(\lambda t+f)^2
\right.\cr &~~~~~~~~~~~~~~~~~~
{}~~~~~~~~~~\left.+i\chi \star D\psi +iD_i\eta\psi^i
-D_i\lambda D^i\phi+{i\over 2}\chi[\chi,\phi]+i[\psi_i,\lambda]\psi^i\right)
.\cr}
\eqn\mmdmdmmd$$
As  before, $H$ can be integrated out, simply setting $H-\lambda t-f=0$.
The benefit from perturbing $L$ to $L(t)$ is that for $t\not= 0$,
$\lambda$, $\chi$, and $\eta$ can also be integrated out, leaving a local
Lagrangian:
$$\eqalign{L'(t)={1\over h^2} & \int_\Sigma \d\mu\Tr\left(
{1\over t}\left(D_ifD^i\phi+if[\psi_i,\psi^i]-iD_l\psi^l\epsilon^{ij}D_i\psi_j
\right)\right.\cr & +\left.{1\over t^2}\left({i\over
2}D_l\psi^l[D_k\psi^k,\phi]
+{1\over 2}\left(-D_kD^k\phi+i[\psi_k,\psi^k]\right)^2\right)\right).\cr}
\eqn\hsh$$
For instance, $\lambda$ is integrated out by setting
$$\lambda=-{f\over t}.      \eqn\tuggo$$

Already we can assert a major point: the standard ``cohomological'' Lagrangian
\mdmdmmd, in which the correlation functions of BRST invariant operators
have a known description in terms of cohomology of ${\cal M}$, can be deformed
preserving the BRST symmetry
to a Lagrangian written in terms of the minimal multiplet $A,\psi,\phi$ only,
namely \hsh.  \mdmdmmd\ and \hsh\ may not be equivalent, but the failure
of such equivalence can only come from new components ${\cal M}_\alpha$
of moduli space that flow in from infinity for $t\not= 0$;
the contribution of the ``old'' component
${\cal M}$ must be independent of $t$.

The terms of order $1/t$ in \hsh\ are
$${i\over t}\{Q,\int_\Sigma \d\mu ~~\Tr\psi^iD_if\}.\eqn\ndnd$$
The terms of order $1/t^2$ are similarly
$${i\over 2t^2}\{Q,\int_\Sigma \d\mu ~~\Tr\left(D_kD^k\phi\cdot D_l\psi^l
-i[\psi_l,\psi^l]D_k\psi^k\right)\}.\eqn\fdfd$$
We will study \hsh\ in the limit of large imaginary $t$,
and since the terms of order $1/t$ already give a nondegenerate kinetic
energy, the terms of order $1/t^2$ can simply be dropped.
Setting $t=-iu$,
we reduce $L'$ to
$$L''(u)={i\over h^2u}\int_\Sigma
\d\mu \Tr\left(D_ifD^i\phi+if[\psi_i,\psi^i]-iD_l\psi^l\epsilon^{ij}D_i\psi_j
\right).              \eqn\hoborgo$$

Now, we want to ask what kind of ``localization'' there is in
the path integral
$${1\over {\rm vol}(G)}\int DA\,\,D\psi\,\,D\phi\,\,\,\,
\exp(-L''(u)).  \eqn\hodo$$
The main point is that the $\phi$ integral is
$$\int D\phi\,\,\exp\left({i\over h^2u}\int_\Sigma \d\mu\,\,
\Tr\phi D_iD^if\right)
\sim \prod_{x\in \Sigma}\delta (D_iD^if). \eqn\odo$$
The localization is therefore on the locus $D_iD^if=0$.
This equation implies
$$0=\int_\Sigma \d\mu\,\Tr fD_iD^if=-\int_\Sigma\Tr (D_if)^2, \eqn\jodod$$
and so it is equivalent to
$$ 0=D_if. \eqn\podod$$
These are the classical Yang-Mills equations, that is, the variational
equations derived from the usual Yang-Mills action $I=-
\int_\Sigma\Tr f^2$.  The space of solutions of \podod\ has one
component, ${\cal M}$, consisting of solutions of $f=0$, and other
components, ${\cal M}_\alpha$, consisting of higher critical points
of the Yang-Mills action.

{}From \tuggo, we see that the new components have $\lambda\sim 1/t$, and
hence are absent at $t=0$ and ``flow in from infinity'' when one perturbs
to $t\not= 0$.  This is the abstract scenario that we anticipated earlier
for how a perturbation of the form $\{Q,\dots\}$ might fail to leave
the theory invariant.

\subsection{Final Reduction}

For any BRST invariant operator ${\cal O}$, let $\langle {\cal O}\rangle$
be the expectation value of ${\cal O}$ computed in the cohomological
theory \mdmdmmd, and let $\langle{\cal O}\rangle'$ be the corresponding
expectation value in the theory \hoborgo.  In general $\langle {\cal O}
\rangle\not= \langle{\cal O}\rangle'$, because of higher critical points
contributing in \hoborgo.  We will describe a class of ${\cal O}$'s such
that the higher critical points do not contribute, and hence
$\langle{\cal O}\rangle=\langle{\cal O}\rangle'$.

Two particular BRST invariant operators will play an important role.
The first, related to the symplectic structure of ${\cal M}$,
is
$$\omega={1\over 4\pi^2}\int_\Sigma\Tr\left(i\phi F+{1\over 2}
\psi\wedge\psi\right).        \eqn\mupo$$
The second is
$$\Theta={1\over 8\pi^2}\int_\Sigma \d\mu \Tr \phi^2.      \eqn\nupo$$

We wish to compute
$$\langle \exp\left(\omega+\epsilon\Theta \right)\cdot\beta\rangle' \eqn\jcc$$
with $\epsilon$ a positive real number, and
$\beta$ an arbitrary observable with at most a polynomial dependence on
$\phi$.
This is
$$\eqalign{\langle\exp\left(\omega+\epsilon\Theta\right)\cdot\beta\rangle'
={1\over {\rm vol}(G)}
\int DA\,\, & D\psi\,\,D\phi ~\cdot\beta\cdot
\exp \left(  {1\over h^2 u}\{Q,\int_\Sigma \d\mu\,\,\psi^kD_kf\}
\right. \cr &\left.
 +{1\over 4\pi^2}\int_\Sigma \Tr\left(i\phi F+{1\over 2}\psi\wedge \psi\right)
 +{\epsilon\over 8\pi^2}\int_\Sigma \d\mu\Tr \phi^2 \right).\cr} \eqn\cupo$$
This is formally independent of $u$, and will really be independent
of $u$ as long as, in varying $u$, the Lagrangian remains nondegenerate
and with a good behavior at infinity in field space.  The particular
choice of $\omega$ and $\Theta$ has been made to ensure that these
conditions are obeyed (for $\epsilon\geq 0$) even at $u=\infty$.
Thus, we can simply
set $u=\infty$ in \cupo, discarding the terms of order $1/u$,
and reducing to
$$\eqalign{
\langle\exp\left(\omega+\epsilon\Theta\right)\cdot \beta \rangle'
={1\over {\rm vol}(G)}
\int DA\,\,D\psi\,\,D\phi \,\,\exp&\left( {1\over 4\pi^2}
\int_\Sigma\Tr\left(i\phi F+{1\over 2}\psi\wedge\psi\right)
\right.\cr&~~~~~~~\left.+{\epsilon\over 8\pi^2}\int_\Sigma
\d\mu\Tr\phi^2\right)
\cdot \beta.\cr} \eqn\kuko$$
This is the key step; we have passed from ``cohomological'' to ``physical''
Yang-Mills theory.

First consider the special case of $\epsilon=0$,
from which all the topological information can be extracted.
If we also set $\beta=1$,
then the
$\phi$ integral gives a multiple of $\prod_{x\in \Sigma}
\delta(F)$, and thus in this special case,
the higher critical points with $F\not= 0$ do not contribute.
Even for $\beta\not=1$ (but still at $\epsilon=0$), the $\phi$ integral
gives a more complicated distribution supported at $F=0$,
\foot{This generalizes the fact that in one dimension,
$f(x)=\int_{-\infty}^\infty \d\phi \,\,\exp(i\phi x)\cdot \phi^n$
is for any positive integer $n$ a distribution supported at $x=0$.}
so still
the higher critical points do not contribute.  As
\hoborgo\ differs from \mdmdmmd\ only by possible contributions of
these higher critical points, the vanishing of these contributions
means that
$$\langle \exp(\omega)\cdot \beta\rangle =\langle \exp(\omega)\cdot\beta
\rangle',
\eqn\basform$$
for all $\beta$ with polynomial $\phi$ dependence.

All information of topological interest can be extracted from \basform.
Knowledge of the left hand side of \basform\ for arbitrary $\beta$
is enough to determine all
expressions $\langle \beta \rangle$ for BRST invariant $\beta$.
(In fact, this is so for a rather elementary reason.  If $\beta$
has definite ghost number, then $\langle\beta\rangle$ vanishes unless
its ghost number is equal to the (real) dimension of ${\cal M}$, and in that
case, since $\omega$ has ghost number 2, $\langle\beta\rangle=\langle
\exp(\omega)\cdot\beta\rangle$.)
On the other
hand,  the right hand side of \basform\ is effectively computable
using \kuko, as will be clear in \S4-5.

The exponent in \kuko\ is a Lagrangian
$$L(A,\psi,\phi)=\int_\Sigma\Tr\left(-i\phi F-{1\over 2}\psi\wedge\psi\right)
-{\epsilon\over 8\pi^2}\int_\Sigma
\d\mu \Tr\phi^2 \eqn\nuko$$
that is entirely equivalent to conventional two-dimensional Yang-Mills
theory.  In fact, $\psi$ is a decoupled field with only a mass term
and Euler-Lagrange equation $\psi=0$; its only role is to put things
in the right theoretical context.  (Including $\psi$ in this way was
originally suggested several years ago by S. Axelrod.)
$\phi$ can also be integrated out, to put the Lagrangian in its conventional
Yang-Mills form.

\subsection{Interpretation Of The Measure}

In quantum gauge theories in general, on a space-time manifold $M$,
the path integral measure on the
space ${\cal A}$ of connections is usually
defined by first introducing a metric on $M$.  This permits one to define
on ${\cal A}$
a metric as follows: a tangent vector to ${\cal A}$ is an adjoint-valued
one-form $a$, and one sets
$$|a|^2=-{1\over 4\pi^2}\int_\Sigma \Tr a\wedge \star
a.\eqn\dndn$$
{}From this metric one formally gets a measure, and this is the usual
path integral measure for gauge fields.

However, in two dimensions, if $M$ is orientable, there is another approach
to defining a measure: this comes from the symplectic structure on
${\cal A}$, that we have noted in \urgo.  Moreover, the measures on
${\cal A}$ defined by the metric or the symplectic structure agree,
because the metric \dndn\ is Kahler.

Look back to \kuko, assuming first that $\beta=1$.
In this case, the only $\psi$ dependent factors are in
$$ DA \,\,D\psi \,\,\,\exp({1\over 4\pi^2}\int_\Sigma\Tr \psi\wedge\psi).
       \eqn\msms$$
This should be compared to the integral in \naturalmeasure.
As in the discussion of that equation, integrating out
$\psi$ will give the symplectic or Liouville measure on ${\cal A}$,
which is the usual path integral measure.
In keeping with convention (but somewhat inconsistently), we will call
this measure $DA$.

Thus, if $\beta=1$, the only role of $\psi$ was to give a more sophisticated
way to build in the standard measure on ${\cal A}$.  Things
are different if $\beta\not=1$ and more specifically if $\beta$ depends on
$\psi$.  In that case, integrating out $\psi$ will replace $\beta$ by
some function of $A$ and $\phi$ only.  This will be a major step in
the detailed computations in \S4 and \S5.

\subsection{Elimination Of $\phi$}

The other main step in the calculations is closely related to the ability
to eliminate $\phi$.
This is possible because derivatives of $\phi$ are absent in
\kuko.  (This is so even if $\beta\not=1$; the BRST cohomology of the
theory can be represented by operators that do not involve derivatives of
$\phi$.)
At this point, let us generalize to $\epsilon\not=0$, but for simplicity
$\beta=1$.
In this case, by integrating out $\phi$ (after integrating out $\psi$
as discussed above), we get
$$\langle\exp\left(\omega+\epsilon\Theta \right)\rangle'
={1\over {\rm vol}(G)}
\int DA\,\,\,\,\exp\left({2\pi^2\over \epsilon}
\int_\Sigma \d\mu\Tr f^2\right). \eqn\nokuko$$
This is the path integral of
conventional two dimensional Yang-Mills theory.  Now, at $\epsilon\not= 0$,
we cannot claim that the $\langle ~~~\rangle$ and $\langle ~~~\rangle'$
operations coincide, since the higher critical components ${\cal M}_\alpha$
contribute.  However, their contributions are exponentially small, involving
the relevant values of $I=-\int_\Sigma \d\mu \Tr f^2$.  So we get
$$\langle\exp\left(\omega+\epsilon\Theta\right)\rangle
={1\over {\rm vol}(G)}
\int DA\,\,\,\,\exp\left({2\pi^2\over\epsilon}
\int_\Sigma \d\mu\Tr f^2\right)
+O(\exp(-2\pi^2c/\epsilon), \eqn\cun$$
where $c$ is the smallest value of the Yang-Mills action $I$ on
one of the higher critical points.

For $\beta\not=1$, the elimination of $\phi$ is more elaborate, but can still
be carried out explicitly. A version of this will be done in our
detailed calculation of intersection numbers in \S5.

\section{The BRST Cohomology}

Now we want to describe the BRST invariant observables of the theory.

Let $T$ be a homogeneous invariant polynomial on the Lie algebra ${\cal H}$.
Then a BRST invariant operator that cannot be written as $\{Q,\dots\}$ is
$$ {\cal O}^{(0)}_T(P)=T(\phi(P)), \eqn\jujitsu$$
with an arbitrary $P\in\Sigma$.  However, for $P,P'\in\Sigma$, the difference
$T(\phi(P))-T(\phi(P'))$ is $\{Q,\dots\}$; this follows from the formula
$$ \d{\cal O}^{(0)}_T=-i\{Q,{\cal O}^{(1)}_T\}, \eqn\umitsu$$
with
$${\cal O}^{(1)}_T=-{\partial T\over\partial \phi^a}\psi^a. \eqn\mitsu$$
One similarly has
$$\d{\cal O}^{(1)}_T=-i\{Q,{\cal O}^{(2)}_T\},
    \eqn\itsu$$
with
$${\cal O}^{(2)}_T= {1\over 2}{\partial^2T\over\partial\phi^a\partial\phi^b}
\psi^a\wedge\psi^b+i{\partial T\over\partial\phi^a}F^a. \eqn\tsu$$
Here $F^a$ are the components of the curvature two-form $F$, that
is $F=\sum_aT_aF^a$.

The following BRST invariant observables can be formed from these
${\cal O}$'s.  Since \itsu\ asserts that ${\cal O}^{(2)}_T$ is annihilated
by $Q$ up to an exact form, we have the $Q$-invariant observable
$$T_{(2)}=\int_\Sigma{\cal O}^{(2)}_T. \eqn\pitsu$$
Likewise, for every oriented circle $C\subset \Sigma$, we have
$$T_{(1)}(C)=\int_C{\cal O}^{(1)}_T. \eqn\ritsu$$
\umitsu\ implies that this is $Q$-invariant, and
\itsu\ implies that, up to $\{Q,\dots\}$, it depends only on the
homology class of $C$.  Finally, we have the original operator
$$T_{(0)}(P)={\cal O}^{(0)}(P). \eqn\citsu$$
To unify the notation, note that for $j=0,1,2$, and $V$ a $j$-dimensional
submanifold of $\Sigma$ (so $V$ is a point for $j=0$, a circle for $j=1$,
and $V=\Sigma$ for $j=2$), we have defined an operator $T_{(j)}(V)$.
Notice that if $A,\psi,\phi$ are considered to have degree or ghost
number $0,1,2$, and $T$ is of order $r$, then $T_{(j)}$ is of ghost number
$2r-j$.

In applications, it may be convenient to replace $T_{(0)}(P)$ by its
averaged version
$$T_{(0)}(P)\to \int_\Sigma \d\mu \,\,\,T(\phi)
\eqn\avver$$
which we will use in \S4-5.  Of course, the two are cohomologous.

\subsection{Mathematical Counterpart}

Now, mathematically, observables with properties analogous to these
can be defined as follows.  The moduli space ${\cal M}$ of flat
connections on an $H$ bundle $E$ parametrizes a family of flat
bundles on $\Sigma$.  One can try to fit them together into a
``universal bundle'' ${\cal E}$ over ${\cal M}\times\Sigma$.
The obstruction to existence of ${\cal E}$ comes from possible
symmetries of flat connections.  If one restricts to a dense open set
in ${\cal M}$ parametrizing irreducible flat connections, the only
symmetries are constant gauge transformations by elements of the center
of $H$.  In the adjoint representation, these act trivially, so the universal
bundle exists at least as an adjoint bundle; this is good enough for
defining the rational characteristic classes that we want.

So pick a connection $B$ on ${\cal E}$, and let ${\cal F}=dB+B\wedge B$
be the curvature.  Then, for $T$ as above,
the closed $2r$-form $T({\cal F})$ defines
an element of $H^{2r}({\cal M}\times\Sigma,\IR)$.  If $V$ is a $j$-dimensional
submanifold of $\Sigma$, then by restricting $T({\cal F})$ to
${\cal M}\times V$, and integrating over the fibers of the projection
${\cal M}\times V\to {\cal M}$, one gets elements
$\widehat T_{(j)}(V)$ in $H^{2r-j}({\cal M},\IR)$.

It is striking that the $\widehat T_{(j)}(V)$
have the same degree $2r-j$
and are determined by the same data as the $T_{(j)}$.
In fact, it can be shown that in ``cohomological gauge theories,''
as long as ${\cal M}$ is non-singular, there is a precise correspondence
between the $T_{(j)}$'s and the $\widehat T_{(j)}$'s, in the sense
that (as long as singularities of ${\cal M}$ can be neglected)
$$\langle\prod_{s=1}^nT_{j_s}(V_s)\rangle={1\over \#Z(H)}\int_{{\cal M}}
\prod_{s=1}^n\widehat T_{(j_s)}(V_s). \eqn\ococ$$
(The factor of $1/\#Z(H)$ is explained in \S2.2 of [\ewitten].)
See [\baulieu] for an explanation of this formula
in the context of four dimensional
Donaldson theory.
The discussion carries over without modification for the analogous
two dimensional theory \mdmdmmd.  Therefore, following through
our derivations above, \ococ\ holds for the contribution of $\mu^{-1}(0)$
if the left hand side of \ococ\ is computed using
the simplified cohomological Lagrangian \hsh,
and holds modulo terms that are exponentially small for $\epsilon\to 0$
if one uses instead the physical Yang-Mills Lagrangian
\nuko.

\subsection{Reduction To Generators}

Elementary arguments show
that in case $T$ can be factored as a product of invariant polynomials,
say $T=UV$, then the
operators $T_{(j)}$ can be expressed in terms of the $U_{(j)}$ and
$V_{(j)}$.
In fact, it is fairly obvious
that up to $\{Q,\dots\}$,
$$T_{(0)}=U_{(0)}V_{(0)},\eqn\firsteqn$$
 and for any circle $C$
$$T_{(1)}(C)=U_{(0)}V_{(1)}(C)+U_{(1)}(C)V_{(0)}.\eqn\secondeqn$$
The corresponding factorization of $T_{(2)}$ is slightly more complicated.
Let $C_\sigma,\,\,\sigma=1\dots 2g$ be circles representing a basis
of $H_1(\Sigma,\IZ)$, with intersection form $\gamma_{\sigma\tau}$.
Then up to $\{Q,\dots\}$,
$$T_{(2)}=U_{(2)}V_{(0)}+U_{(0)}V_{(2)}+\sum_{\sigma,\tau}
\gamma_{\sigma\tau}U_{(1)}(C_\sigma)V_{(1)}(C_\tau).\eqn\thirdeqn$$
These formulas show that it is sufficient to evaluate \ococ\ with
the $T$'s taken from a set of generators of the ring of invariant
polynomials on ${\cal H}$.

\chapter{Localization And Yang-Mills Theory}

Our goal in this section is to make a detailed comparison
of the localization theorem with two dimensional Yang-Mills theory.
We will also begin the computation of intersection numbers on
moduli spaces of flat connections; these computations will be completed
in the next section.

To begin with, we need some basic facts about the quantum gauge theory.
I will here explain these facts from a continuum point of view.
This discussion can be read in conjunction with \S2.3 of [\ewitten]
where many of the same facts are derived using a lattice regularization,
with more detail on some points.  Other approaches have been cited in the
introduction.

Notation is generally as in earlier sections.
Thus, $A$ is a connection on an $H$-bundle $E$ over a Riemann surface $\Sigma$.
$F=\d A+A\wedge A$ is the curvature.  $\phi$ is a zero-form with values in
$\ade$.  We write $\phi=\sum_a\phi^aT_a$, with $T_a$ an orthonormal basis
of the Lie algebra ${\cal H}$ of $H$.  At the outset, we assume that $H$
is connected and simply connected.  Later we relax the assumption of simple
connectivity.

\subsection{The Topological Field Theory}

First we consider the topological field theory with Lagrangian
$$L = -{i\over 4\pi^2}\int_\Sigma\Tr \phi F, \eqn\toplag$$
which is related to Reidemeister-Ray-Singer torsion [\schwarz].
The partition function is defined formally by
$$Z(\Sigma)={1\over {\rm Vol}(G)}\int DA\,\,\,D\phi \,\,\,\,\,
\exp(-L).\eqn\ufgo$$
Here if $E$ is trivial, $G$ is the group of maps of $\Sigma$ to $H$;
in general $G$ is the group of gauge transformations.

\toplag\ should really
be considered to correspond to a one parameter family of topological
field theories.  Different methods of defining the path integral will
differ by terms coming from a substitution
$$L\to L+v\int_\Sigma\d\mu\,\,{R\over 4\pi}, \eqn\oplag$$
where $v$ is an arbitrary constant,
and $R$ is the curvature of a metric on $\Sigma$ (which might enter
in regularizing and gauge fixing the theory; or if one uses a lattice
regularization as in [\ewitten], a similar ambiguity arises in defining
the local factors).  Of course, the substitution \oplag\ just multiplies
the path integral on a surface $\Sigma$ of Euler characteristic $\chi(\Sigma)$
by a ``trivial'' factor $\exp(-v\chi(\Sigma))$.
We ultimately will fix $v$ by requiring precise agreement (and not just
agreement
up to such a trivial factor) with the theory of Reidemeister-Ray-Singer
torsion.
In principle, with a careful calculation using a regulator in which the
manipulations of \S3 are valid, one should hopefully be able to determine
{\it a priori} the necessary value of $v$.

\subsection{Canonical Quantization}

\def\wph{\widetilde\phi}
Canonical quantization of \toplag\ shows that the canonical momentum to $A$ is
$$ \pi_A=\wph,           \eqn\goglag$$
where we set $\wph=\phi/4\pi^2$.
Hence in the quantum theory, if the $A^a$ are taken to be multiplication
operators, then  $\wph$ acts as follows:
$$ \wph_a=-i{\delta \over\delta A^a(x)}. \eqn\oglag$$

\def\al{\alpha}

If $C$ is a circle in $\Sigma$ -- an initial value surface -- then the
Hilbert space ${\cal H}_C$ obtained by quantization on $\Sigma$ can be
considered with this representation of the canonical commutation relations
to consist of gauge invariant functions $\Psi(A)$.  Such a function must
be a function only of the monodromy of $A$ around $C$, which (picking
a base point $x\in C$) we write as $$ U=P\exp\oint_x A. \eqn\bubn$$
To be more precise, $\Psi$ must be a class function of $U$, invariant
under conjugation, and so must have an expansion in characters:
$$ \Psi(A)=\sum_\al c_\al\,\,\,\chi_\al(U). \eqn\ubn$$
The sum runs over all isomorphism classes of irreducible $H$ representation
$\al$;
$\chi_\al(U)$ is the trace of $U$ in the representation $\al$;
$c_\al$ are complex numbers.  Thus the functions $\chi_\al(U)$ give a
basis
of ${\cal H}_C$.  The lattice regularization used in [\ewitten]
makes it clear that this basis is orthonormal.

\subsection{Casimir Operators}

Let $Q$ be an invariant polynomial on ${\cal H}$.
Pick $y\in \Sigma$.  We want to determine the quantum operator $\widehat Q$
on ${\cal H}_C$ corresponding to the classical observable $Q(\widetilde\phi)$.

The result can be determined from \oglag, apart from a normal ordering
ambiguity that will be treated later.  We have
$$\wph_a\cdot\chi_\al(U)=-i\Tr_\al T_a P\exp\oint_y A\eqn\pibo$$
since the right hand side is the first variation of $\chi_\al(U)$ with
respect to $A$.  Formally computing
higher derivatives in the same way, we get
$$ Q(\wph)\to \widehat Q =Q(-iT). \eqn\ibo$$
The right hand side is just the Casimir operator determined by $Q$.

\ibo\ is good enough for our purposes temporarily, but it has the following
limitation.  If one studies the theory \toplag\ with an arbitrary
regularization, then gauge invariant operators such as $Q(\widetilde\phi)$
will ``mix'' under renormalization -- by a normal ordering ambiguity -- with
similar operators determined by lower order polynomials. \ibo\ is one
natural choice, but we will ultimately have to modify it.
For any given gauge group, there are only finitely many independent
Casimir operators, and therefore the renormalization problem involves
finitely many parameters.  For instance, for $SU(2)$ or $SO(3)$, one such
parameter will appear.

\subsection{The Three-Holed Sphere}

\FIG\three{(a) A three-holed sphere, with the boundary components labeled
by representations $\alpha_i$.  (b) The same path integral with an insertion
of an operator ${\cal O}$. (c) One can consider this operator to act on any
of the three external states.}
\FIG\two{A two-holed sphere with labeled boundaries.}
\FIG\one{A one-holed sphere with a labeled boundary.}
Every oriented Riemann surface can be built by gluing together
three-holed spheres, so the path integral on a three-holed sphere $\Sigma$
(figure (\three))  is an important special case.
Let $C_i,\,\,\,\,i=1\dots 3$ be the three
boundary circles. Let $U_i$ be the holomony of the connection
$A$ about $C_i$ (orientations on the $C_i$, used in computing the holonomy,
are induced from $\Sigma$).
The path integral on the three-holed sphere
gives a vector $\Psi_{\{3\}}\in \otimes_{i=1}^3
{\cal H}_{C_i}$, which must therefore
be of the form
$$\Psi_{\{3\}}=\sum_{\al_1,\al_2,\al_3}c_{\al_1,\al_2,\al_3}
\prod_{i=1}^3\chi_{\al_i}(C_i).
\eqn\olmo$$

\REF\zel{D. P. Zelobenko, ``Compact Lie Groups And Their Representations,''
Translations of Mathematical Monographs, vol. 40 (American Mathematical
Society, 1978).}
I now claim, however, that $c_{\al_1,\al_2,\al_3}$ vanishes unless $\al_1
=\al_2=\al_3$.
To see this, we consider the path integral of figure (\three(b)) with
the insertion of some operator ${\cal O}=
Q(\wph)$ at some point $z\in \Sigma$.
By factorizing in an appropriate channel, as in figure (\three(c)), one can
consider this operator to act on any of the three external states $\chi_{\al_i}
(U_i)$, whence, according to \ibo, ${\cal O}$ can be replaced by the value of
the Casimir operator $Q(-iT)$ for the representation $\al_i$.  Hence
$c_{\al_1,\al_2,\al_3}=0$ unless the $\al_i$
have the same values for all Casimirs.
This implies that they must be isomorphic [\zel, \S126].
\olmo\ thus collapses to
$$\Psi_{\{3\}}=\sum_\al c_\al\prod_{i=1}^3\chi_\al(U_i). \eqn\oco$$

\subsection{The Two-Holed Sphere}

We now consider the two-holed sphere of figure (\two), with boundary components
$C_1,C_2$ and monodromies about them $U_1,U_2$.  Labeling the
boundary components by representations $\al_i$, the path
integral gives a result of the general form $\Psi_{\{2\}}=\sum_{\al_1,\al_2}
e_{\al_1,\al_2}\chi_{\al_1}(U_1)\chi_{\al_2}(U_2)$.
However, $e_{\al_1,\al_2}=0$ unless
$\al_1=\al_2$, by the argument in the last paragraph, so we can write
$e_{\al_1,\al_2}=\delta_{\al_1,\al_2}e_{\al_1}$, for some $e_\al$'s.

The path integral of the two-holed sphere with labeled external
states can be given a special interpretation.  In fact,
$e_{\al_1,\al_2}=\langle
\chi_{\al_2}|
\exp(-HT)|\chi_{\al_1}\rangle$, where $H$ is the Hamiltonian and
$T$ is the elapsed time.  As $H=0$ for the topological field theory
\toplag, we have merely $e_{\al_1,\al_2}=\delta_{\alpha_1,\alpha_2}$,
so $e_\al=1$ and
$$\Psi_{\{2\}} =\sum_\al\chi_\al(U_1)\chi_\al(U_2).      \eqn\ococo$$

\subsection{The One-Holed Sphere}

The path integral of the disc or one-holed sphere (figure (\one))
similarly gives a result of the general form
$$\Psi_{\{1\}}(U)=\sum_\al f_\al\chi_\al(U)        \eqn\xococox$$
with $U$ the monodromy about the boundary.
Now, however, we can reason as follows.
Writing down the path integral from \toplag\
$$\int DA\,\,\,D\phi\,\,\,\,\,\,\exp\left({i\over 4\pi^2}\int_\Sigma
\Tr\phi F\right).
    \eqn\zozoz$$
we see that the $\phi$ integral gives explicitly  $\delta(F)$.

A connection on the disc with $F=0$ necessarily has $U=1$, so the function
$\Psi_{\{1\}}(U)$ in \xococox\ must be a delta function supported at $U=1$.
According to the theory of compact Lie groups, we have
$\delta(U-1)=\sum_\al \dim(\al)\chi_\al(U)$, with $\dim(\al)$
the dimension of the representation $\al$.  Thus
$$f_\al= w\cdot \dim(\al), \eqn\ozoz$$
with $w$ an unknown constant whose origin was explained
in connection with \oplag.

\FIG\factoring{Factoring the amplitude of a two-holed sphere in
a suitable fashion.}
We can now identify the unknown constants $c_\al$ in the path integral
on a three-holed sphere.  As in figure (\factoring), we decompose a two-holed
sphere as the union of a one-holed sphere and a three-holed sphere, glued on
their common boundary.  Factorizing the partition function of the three-holed
sphere, we
learn that
$$ e_\al=c_\al f_\al. \eqn\zoz$$
Since we already know $e_\al=1$, we get
$$ c_\al={w^{-1}\over \dim(\al)}. \eqn\xoxozo$$
For a derivation of these results using a lattice regularization,
see [\ewitten,\S2.3].

\subsection{Combining The Pieces}
\FIG\cutpaste{A surface of genus $g$ -- in this case $g=2$ -- can
be regarded as a union of $2g-2$ three-holed spheres.}
Now we can combine the pieces and determine the partition function of
the topological field theory \toplag\ on an oriented Riemann surface
$\Sigma$ of genus $g$.  Such a surface can be regarded, as in figure
(\cutpaste), as the union of $2g-2$ three-holed spheres, glued on $3g-3$
circles.  Labeling each circle $C_i$ by a representation $\alpha_i$,
and computing the path integral on each three holed sphere using \oco,
one finds that the sum over the $\alpha_i$ collapses to a single sum, because
of the ``diagonal'' nature of the partition function of the three-holed sphere.
The result
for the partition function is thus $Z(\Sigma)=\sum_\al c_\al^{2g-2}$.

After adjusting $w$ by comparing to Reidemeister-Ray-Singer torsion,
as explained in [\ewitten],
the partition function is finally
$$Z(\Sigma)= \left({{\rm Vol}(H)\over (2\pi)^{\rm dim H}}\right)^{2g-2}
\cdot \sum_\al{1\over (\dim \al)^{2g-2}}. \eqn\moxox$$

The moduli space ${\cal M}$ of flat connections on $\Sigma$ has
the symplectic structure $\omega$ of equation \urgo, and hence
a volume form $\omega^n/n!$
As explained in [\ewitten, \S2], the partition function $Z$ is related
to the volume of ${\cal M}$ by $Z={\rm Vol}({\cal M})/\# Z(H)$.
(One divides here by $\# Z(H)$, which is the number of elements of the center
of $H$,  essentially
because the generic connection has $\# Z(H)$ symmetries.)
Hence the volume of ${\cal M}$ is
$$\int_{{\cal M}}\exp(\omega)=\int_{{\cal M}}{\omega^n\over n!}
=\#Z(H)\cdot\left({{\rm Vol}(H)\over (2\pi)^{\rm dim H}}\right)^{2g-2}
\cdot \sum_\al{1\over (\dim \al)^{2g-2}}. \eqn\omoxox$$
This is then a first special case of evaluation of
the intersection pairings on moduli space.

\section{Groups With Non-Trivial $\pi_1$}

So far we have assumed that the gauge group $H$ is connected and
simply connected.  These conditions ensure that an $H$ bundle $E$ over
a two dimensional surface $\Sigma$ is trivial.  We now wish to drop
the condition of simple connectivity.

Let $\Gamma$ be a subgroup of the center $Z(H)$.  We take the gauge
group to be $H'=H/\Gamma$, which of course is still connected but is not
simply connected.

\subsection{Classification Of Bundles}

An $H$ bundle on a two dimensional surface is necessarily
trivial, but this is not so for $H'$ bundles.  The possible $H'$ bundles
have the following standard description.  Let $\Sigma$ be a closed
oriented Riemann surface of genus $g$.
Let $E'$
be a principal $H'$ bundle over $\Sigma$.  Let $P$ be a point in $\Sigma$.
$E'$ is necessarily trivial when restricted to $\Sigma-P$ (that is,
$\Sigma$ with $P$ deleted).  Thus, on $\Sigma-P$, $E'$ can be lifted
to a principal $H$ bundle $E$.

A connection $A'$ on $E'$, when restricted to $\Sigma-P$, lifts to a
connection $A$ on $E$.  However, $A$ does not necessarily extend smoothly
over $P$.  The monodromy $u$ of $A$ about $P$ is an element of $H$ that
projects to the identity in $H'$ (since $A'$ extends smoothly over $P$).
Thus, $u$ is an element of $\Gamma$.  It is easy to see that $u$ is
a topological invariant of $E'$.
Conversely, it is standard that $u$ is the only such invariant and
can take arbitrary values.  Thus, the possible $H'$ bundles $E'(u)$ are
classified by the arbitrary choice of $u\in \Gamma$.

\subsection{The Non-Singular Cases}

We will develop most of the story for general $H'$ and $E'(u)$.
However, the topological results that arise are easiest to understand
in cases in which the space of flat connections
on $E'(u)$ is smooth, and is acted on freely by the gauge group.
This implies in particular that the moduli space ${\cal M}'(u)$ of flat
connections on $E'(u)$ is smooth.

The main case in which this occurs is the following.
Let $\Sigma$ be a Riemann surface of genus $\geq 1$.
Let $H=SU(N)$, and let $H'=H/\Gamma$, with $\Gamma\cong \IZ/N\IZ$
the center of $SU(N)$.  Finally, let $u$ be a generator of $\Gamma$.
In such cases, the topological conditions that we want are known to hold.
(For $g=1$, the moduli space of flat connections on $E'(u)$
is a single point, for $g>1$ a smooth closed
manifold of dimension $(N^2-1)\cdot (2g-2)$.)
The simplest example, which we use later for illustration,
is $H=SU(2)$, $H'=SO(3)$, $u=-1$.

\subsection{The Volume Of The Gauge Group}

We want to perform the $H'$ path integral
$${1\over {\rm Vol}(G')}\int DA'\,\,D\phi\,\,\,\,\,\exp(-L), \eqn\moxoxzz$$
with $G'$ the group of gauge transformations of $E'(u)$.
We will do this by relating the integral over $A'$ to an integral over its
lift $A$.  We must worry about two points: the comparison of volumes
of $G'$ and $G$, which we consider first; and the role of the singularity
at $P$.

The group $G$ of maps of $\Sigma$ to the connected
and simply connected group $H$ is connected.  This is not true for
$G'$; the set of components of $G'$ can be identified with
the finite group ${\rm Hom}(H_1(\Sigma,\IZ),\Gamma)$.  This group
has $\# \Gamma^{2g}$ elements.   Thus, if $G'_1$ is the identity
component of $G'$, we have ${\rm Vol}(G')=\# \Gamma^{2g}{\rm Vol}(G'_1)$.

On the other hand, applying pointwise
the projection $H\to H'=H/\Gamma$ gives a natural
map $G\to G'_1$ whose kernel consists of constant gauge transformations
by elements of $\Gamma$.  Hence ${\rm Vol}(G)=\# \Gamma\cdot {\rm Vol}(G'_1)$.

Combining these formulas, we have
$${\rm Vol}(G)=\# \Gamma^{1-2g}{\rm Vol}(G').  \eqn\jjj$$
For future use, we also note the following elementary facts:
$$\eqalign{ {\rm Vol}(H) & = \# \Gamma \cdot {\rm Vol}(H') \cr
            \# Z(H) & = \# \Gamma \cdot \# Z(H')
\cr             \# \pi_1(H')& = \#\Gamma . \cr} \eqn\kkk$$

\subsection{The Singularity At $P$}
\FIG\disccon{Factorizing the genus $g$ amplitude on a disc $D$ containing
the point $P$.}

We now come to the essence of the matter, which is the role of the singularity
at $P$.
This can be deduced by cutting out of $\Sigma$ a disc $D$ containing
$P$ -- so as to factorize the computation on the Hilbert space ${\cal H}_C$
associated with $C=\partial D$.
The path integral over $\phi$ is still localized -- as in \zozoz\ -- on
connections with $F=0$ away from $P$.  Because the monodromy about $P$ is
prescribed to
be $u\in \Gamma$, the monodromy $U$ around $C$ is $u$ rather than 1.
In the derivation of \ozoz, we must replace $\delta(U-1)=\sum_\al\dim(\al)
\chi_\al(U)$ by $\delta(U-u)=\sum_\al\dim(\al)\chi_\al(U)\cdot \lambda_\al
(u^{-1})$.
Here $\lambda_\al(u^{-1})$
is the following.  In the $\al$ representation of $H$,
the element $u^{-1}$ of the center of $H$ is represented by a complex
number of modulus one that we have called $\lambda_\al(u^{-1})$.
(So $\lambda_\al(u^{-1})=\chi_\al(u^{-1})/\dim(\al)$.)

Thus, the role of the singularity at $P$ is to bring about
a substitution
$$\sum_\al\dim(\al)\chi_\al(U)\to \sum_\al\dim(\al)
\chi_\al(U)\cdot \lambda_\al(u^{-1}).
    \eqn\mcmcmc$$
Upon gluing the disc $D$ into the rest of $\Sigma$ and carrying
out the overall evaluation of the path integral, the only role of
the singularity is to multiply the contribution of a representation
$\al$ by a factor of $\lambda_\al(u^{-1})$.
(In [\ewitten,\S3], the factor of $\lambda_\al(u^{-1})$ is extracted
in the special case $H=SU(2)$, $H'=SO(3)$, $u=-1$ from the Verlinde
formula.  This can presumably be done in general.)

\subsection{Evaluation Of The Twisted Partition Function}

Now we want to calculate the $H'$ partition function
$$\ZZ(\Sigma;u)={1\over {\rm Vol}(G')}\int DA'\,\,D\phi\,\,\,
 \exp(-L).\eqn\jico$$

First we calculate the corresponding $H$ partition function for
connections on $\Sigma-P$ with monodromy $u$ around $P$.  This is
$$Z(\Sigma;u)={1\over {\rm Vol}(G)}\int DA\,\,D\phi
\,\,\, \exp(-L). \eqn\ico$$
$A$ is the lift of $A'$.
{}From what we have just said, this is given by the same formula as \moxox\
but weighting each representation by an extra factor of $\lambda_\al(u^{-1})$.
So
$$Z(\Sigma;u)= \left({{\rm Vol}(H)\over (2\pi)^{\rm dim H}}\right)^{2g-2}
\cdot \sum_\al{\lambda_\al(u^{-1})\over (\dim \al)^{2g-2}}. \eqn\moxoxm$$

We now use \jjj\ to relate $Z(\Sigma;u)$ to $\ZZ(\Sigma;u)$, and also
\kkk\ to express the result directly in terms of properties of $H'$.
Using also \kkk, we get
$$\ZZ(\Sigma;u)= {1\over \# \pi_1(H')}
\left({{\rm Vol}(H')\over (2\pi)^{\rm dim H'}}\right)^{2g-2}
\cdot \sum_\al{\lambda_\al(u^{-1})\over (\dim \al)^{2g-2}}. \eqn\mooxox$$
Note that in this formula, the sum runs over all isomorphism classes
of irreducible representations of the universal cover $H$ of $H'$.

Here is a check.  Note that $v\to \lambda_\al(v)$
is a character of the finite abelian group $\Gamma$.
Irreducible $H'$ modules are the same as irreducible $H$ modules for
which this character is trivial.  By the orthogonality of the characters,
$\sum_{u\in \Gamma}\lambda_\al(u)$ vanishes unless $\lambda_\al$ is trivial,
in which case of course it equals $\# \Gamma=\#\pi_1(H')$.
Hence we get
$$\sum_{u\in\gamma}\ZZ(\Sigma;u)=\left({\rm Vol}(H')\over (2\pi)^{{\rm dim}
H'}\right)^{2g-2}{\sum_\al}'{1\over (\dim \al)^{2g-2}}. \eqn\longway$$
Here $\sum'$ is a sum over isomorphism classes of irreducible $H'$ modules.
On the left
the sum over $u\in\Gamma$ should be interpreted as a sum over
all isomorphism classes of $H'$ bundles.  In general, if $u$ is construed
to label these isomorphism classes, \longway\ is true,
as explained in [\ewitten,\S2.3], even if $H'$ is not connected.
If $H'$ is connected and simply connected,
there is only one isomorphism class of $H'$ bundle on $\Sigma$, and \longway\
reduces back to \moxox.

Just as in our discussion of simply connected gauge groups,
$\ZZ(\Sigma;u)$ can be interpreted as $1/\# Z(H')$ times the volume
of the moduli space ${\cal M}'(u)$ of flat connections on the bundle
$E'(u)$.   ($\# Z(H')$ arises, again, as the number of symmetries of
a generic $H'$ connection $A'$.)  So the volume of ${\cal M}'(u)$ is
$$\int_{{\cal M}'(u)}\exp(\omega)=
  {\# Z(H')\over \# \pi_1(H')}
\cdot\left({{\rm Vol}(H')\over (2\pi)^{\rm dim H'}}\right)^{2g-2}
\cdot \sum_\al{\lambda_\al(u^{-1})\over (\dim \al)^{2g-2}}. \eqn\moomox$$

\section{Physical Yang-Mills Theory}

Now we leave the topological field theory \toplag,
and turn to physical Yang-Mills theory.
We introduce on $\Sigma$ a measure $\d\mu$ of total measure 1, and
consider
$$L = -{i\over 4\pi^2}\int_\Sigma\Tr \phi F-{\epsilon\over 8\pi^2}
\int_\Sigma \d\mu\Tr \phi^2, \eqn\physlag$$
with $\epsilon$ a positive real number.
We wish to evaluate the corresponding path integral
$$\int DA \,\,D\phi \,\,\,\exp\left({i\over 4\pi^2}\int_\Sigma
\Tr\phi F+{\epsilon\over 8\pi^2}\int_\Sigma \d\mu \Tr \phi^2\right).
\eqn\nxnxnx$$
Upon performing the Gaussian integral over $\phi$ (or eliminating $\phi$
by its classical equations of motion),  we see that the same
theory could be defined by the Lagrangian
$$I=-{1\over 2\epsilon'}\int_\Sigma \d\mu \Tr f^2, \eqn\otherlag$$
with $f=\star F$ and $\epsilon'=4\pi^2\epsilon$.
(The Yang-Mills Lagrangian is most often written in terms
of $\epsilon'$, but the topological formulas are perhaps most naturally
written in terms of $\epsilon$.)
For $\epsilon\not=0$, the full diffeomorphism invariance
of the topological field theory that we have discussed up to this
point is reduced to
invariance under the group of area-preserving diffeomorphisms.

We should now make a preliminary observation, analogous to the remark
following \ufgo.  Different recipes for defining the path integral
in \physlag\ will differ (in addition to the ambiguity already cited in
\oplag) by
$$L\to L+ t\epsilon'\int_\Sigma \d\mu, \eqn\therlag$$
with $t$ an arbitrary parameter.
This term respects the invariance under area-preserving diffeomorphisms.
Just as we fixed the ambiguity noted
in \oplag\ to agree with the theory of Reidemeister-Ray-Singer torsion,
we will at a judicious moment adjust the value of $t$ to agree with
the topological theory to which we wish to compare.  In principle, a suitable
calculation using a regulator in which the manipulations of \S3 are valid
could probably be used to give an {\it a priori} computation of $t$.
In any case, the extra term in \therlag\ has a ``trivial'' effect on
the partition function, multiplying it by $\exp(-t\epsilon')$.

\subsection{The Classical Solutions}

As we have recalled in the introduction, the space ${\cal A}$ of connections
can be regarded as a symplectic manifold, acted on symplectically by
the group $G$ of gauge transformations; and $I$ can be interpreted as the
norm squared
of the moment map $\mu$,
with respect to a certain invariant quadratic form.
The critical points of $I$ are the classical solutions of two
dimensional Yang-Mills theory.  They have the following explicit description
[\abot].
The Euler-Lagrange equation derived
by varying $I$ is
$$ 0 = Df, \eqn\mxcx$$
with $D$ the gauge-covariant extension of the exterior derivative.
This is certainly obeyed for $f=0$ -- which corresponds to the zeros
of the moment map and the absolute minimum of $I$.   Higher critical
points correspond to $f\not= 0$, in which case $f$, being covariantly
constant, gives a reduction of the structure group of the connection
to a subgroup $H_0$ that commutes with $f$.
Solutions of \mxcx\ can therefore be described rather explicitly:
they are flat $H_0$ connections twisted by
constant curvature line bundles in the $U(1)$ subgroup generated by $f$.

However, to simplify things, in this part of our story we will just consider
the case that the gauge group  is $SU(2)$ or $SO(3)$.  Then $f$, if not
zero, reduces the structure group precisely to $U(1)$, so in the $SU(2)$
case, for instance, we get an $SU(2)$ bundle with a covariantly constant
splitting as a sum of line
bundles.  From the classification of line bundles, it follows at once
that the conjugacy class of $f$ is given by
$$ f=2\pi  m \left(\matrix{i & 0 \cr 0 & -i \cr}\right), \eqn\fromclass$$
with $m\in \IZ$.  The value of $I$ at such a critical point is then
directly computed to be
$$I_m= {(2\pi m)^2\over \epsilon'}.        \eqn\critval$$

In the $SO(3)$ case, there are two isomorphism classes of bundle.
The bundles that lift to $SU(2)$ bundles give the same result
just described.  The non-trivial $SO(3)$ bundles can be described as
$SU(2)$ bundles on $\Sigma- P$ with monodromy $u=-1$ about $P$.  Allowing
for this, \fromclass\ is just replaced by
$$ f=2\pi  (m+1/2) \left(\matrix{i & 0 \cr 0 & -i \cr}\right), \eqn\fromclass$$
with $m\in \IZ$.
The values of $I$ at these critical points are now
$$I'_m={\left(2\pi(m+1/2)\right)^2\over \epsilon'}. \eqn\otherval$$

\subsection{The Hamiltonian}

Our next goal is to evaluate the partition function of physical
Yang-Mills theory, with Lagrangian \physlag, on a Riemann surface of genus
$g$.

The main thing that we need is to evaluate the Hamiltonian.  To this
aim, we pick an initial value circle $C\subset \Sigma$.  We write
the volume form of $\Sigma$
in a neighborhood of $C$ as $d\sigma\wedge d\tau$
where $C$ is defined by $\tau=0$, and $\sigma$ is a periodic parameter
on $C$, with $\oint_C d\sigma = 1$.  The Hamiltonian operator $H$
will be the generator of translations in $\tau$.

Standard canonical quantization shows that
$$H =-{\epsilon'\over 2}\oint \d\sigma \Tr\wph^2
 + t\epsilon'. \eqn\bcbcbc$$
We have included the extra term from \therlag, which just adds a constant
to $H$.

In view of \ibo, this can be described as follows.  Let $C_2$ be the
quadratic Casimir operator $C_2=-\sum_a T_a^2$, with $T_a$ an orthonormal
basis of ${\cal H}$.  Then
$$ H = {\epsilon'\over 2}C_2 + t\epsilon'. \eqn\nuuf$$

When $\epsilon\not=0$, a measure $\mu$
must be introduced on every Riemann
surface we consider.  The simplest case is the two-holed sphere $\Sigma$ of
figure (\two).  Suppose that the total measure is $\rho$.  Then
the area form of $\Sigma $ can be represented by the two form
$d\sigma\wedge \d\tau$, with $\sigma$ as above and $0\leq \tau\leq \rho$.
The path integral on $\Sigma$ can be computed as the matrix element
of $\exp(-\rho H)$ between external states on the boundary.
The generalization of \ococo\ to $\epsilon\not= 0$ is therefore  simply
$$\Psi_{\{2\}}=\sum_\al\chi_\al(U_1)\chi_\al(U_2)\exp\left(
-\epsilon'\rho\left({C_2(\al)\over 2}+t\right)\right).
 \eqn\rococo$$

The $\rho $ dependence of the amplitude on any Riemann surface with an
external line labeled by a representation $\al$ is given by the same
factor $\exp\left(-\epsilon'\rho\left({C_2(\al)\over 2}+t\right)\right)$
as in \rococo, since
one can always increase the area of $\Sigma $ by gluing a cylinder
on to one of the external lines.  Sewing together the external lines,
it follows that also for a Riemann surface without boundary, the contribution
of any representation $\al$ to the partition function has this universal
$\rho$ dependence.

Therefore, we can immediately write down the partition
function, with gauge group $H'$, for connections on a bundle
$E'(u)$, generalizing \mooxox\ to $\epsilon\not=0$.  We get
$$ \ZZ(\Sigma,\epsilon;u)=
{1 \over \# \pi_1(H')}\cdot
\left({{\rm Vol}(H')\over (2\pi)^{\rm dim H'}}\right)^{2g-2}
\cdot \sum_\al{\lambda_\al(u^{-1})\cdot\exp\left(-\epsilon'\left(
{C_2(\al)\over 2}
+t\right)\right)
\over (\dim \al)^{2g-2}}. \eqn\ooxox$$
There is still one unknown parameter $t$.

\section{Comparison With The Localization Formula}

Finally, we can begin to enjoy the fruits of our labors -- comparing
\ooxox\ to the predictions of the localization formula of \S2-3.
We will do this in full detail -- identifying the contributions of higher
critical points -- only for $SU(2)$ and $SO(3)$.  For other groups,
we will study only the contribution of $\mu^{-1}(0)$.

Up to isomorphism, $SU(2)$ has one irreducible representation $\alpha_n$
of dimension
$n$ for every positive integer $n$.
The value of the quadratic Casimir for this
representation is with our normalization $C_2(\al_n) = (n^2-1)/2$.
We set $t=1/4$, so that the eigenvalues of the Hamiltonian are just
$\epsilon' n^2/4$.  As will be clear, this value is required for
agreeing with the predictions of the topological theory.
(Otherwise, the contribution of $\mu^{-1}(0)$ is not a polynomial
in $\epsilon$ but has an extra exponential factor.
Later, we will generalize this determination of $t$ to general compact
Lie groups and general Casimirs.)

First we consider the case of $H=SU(2)$.  Then ${\rm Vol}(SU(2))
=2^{5/2}\pi^2$ with our conventions, and so
$$Z(\Sigma,\epsilon)=  {1\over (2\pi^2)^{g-1}}
\sum_{n=1}^\infty{\exp
\left(-{\epsilon' n^2/4}\right)\over n^{2g-2}}. \eqn\sutwo$$
On the other hand, for a non-trivial $SO(3)$ bundle with $u=-1$,
we have $\lambda_{n}(u^{-1})=(-1)^{n+1}$, $\#\pi_1(H')=2$
and ${\rm Vol}(SO(3))=2^{3/2}\pi^2$, so
$$\ZZ(\Sigma,\epsilon;-1)={1\over 2\cdot (8\pi^2)^{g-1}}\sum_{n=1}^{\infty}
{(-1)^{n+1}\exp(-\epsilon' n^2/4)\over n^{2g-2}}. \eqn\sothree$$

\subsection{Sum Over Critical Points}

We will now show how \sothree\ and \sutwo\ can be written as a sum
over critical points.  In doing so, we consider first the case of genus
$g\geq 1$; then we will return to special features of $g=0$.

First we consider the case of a non-trivial $SO(3)$ bundle.
It is convenient to look at not $\ZZ$ but
$${\partial^{g-1}\ZZ\over \partial\epsilon'{}^{g-1}}
={(-1)^g\over 2\cdot (32\pi^2)^{g-1}}\sum_{n=1}^\infty{(-1)^n\exp(-\epsilon'
n^2/4)}. \eqn\mpp$$
We write
$$\sum_{n=1}^\infty (-1)^n\exp(-\epsilon' n^2/4)
=-{1\over 2}+{1\over 2}\sum_{n\in \IZ}(-1)^n\exp(-\epsilon' n^2/4).
\eqn\ppm$$
The sum on the right hand side of \ppm\ is a theta function,
and in the standard way we can use the Poisson summation formula to
derive the Jacobi inversion formula:
$$\eqalign{\sum_{n\in \IZ}(-1)^n\exp(-\epsilon' n^2/4)
&=\sum_{m\in \IZ}~~~\int_{-\infty}^\infty \d n \exp\left(2\pi i n m +i\pi n
-\epsilon' n^2/4\right)
\cr & = \sqrt{{4\pi\over \epsilon'}}\sum_{m\in
\IZ}\exp\left(-{\left(2\pi(m+1/2)
\right)^2
\over \epsilon'}\right).\cr}     \eqn\icoco$$
Putting the pieces together,
$${\partial^{g-1}\ZZ\over \partial\epsilon'{}^{g-1}}
={(-1)^g\over 4\cdot (32\pi^2)^{g-1}}
\cdot\left(-1+\sqrt{{4\pi\over\epsilon'}}\sum_{m\in \IZ}
\exp\left(-{\left(2\pi(m+1/2)\right)^2
\over \epsilon'}\right)\right).   \eqn\coco$$

Now let us compare this formula to the topological theory explained
in \S2.  For a  non-trivial $SO(3)$ bundle,  $\mu^{-1}(0)$
is smooth and acted on freely by $G$, so we can apply the reasoning of
\S2.4.  For $\epsilon\to 0$, the function $\ZZ(\epsilon)$
should be the sum of a polynomial in $\epsilon$
of degree at most $\dim{\cal M}/4$, plus exponentially small contributions
of unstable critical points.  Looking back to \otherval, where we determined
the values of the square of the moment map at the unstable critical points,
the $m^{th}$ critical point should make a contribution proportional
for small $\epsilon$
to $\exp\left(-\left(2\pi(m+1/2)\right)^2/\epsilon'\right) $, up to a power
of $\epsilon$.  These are precisely the exponents on the right hand side of
\coco.   (Integrating $g-1$ times with respect to $\epsilon$ to recover
$\ZZ(\epsilon)$ from \coco\ will not change these exponents.)

\coco\ shows that $\partial^{g-1}\ZZ/\partial \epsilon'{}^{g-1}$ is a constant
up to exponentially small terms, and hence $\ZZ(\epsilon)$ is a polynomial
of degree $g-1$ up to exponentially small terms.
The terms of order $\epsilon^k,\,\,\,k\leq g-2$ that have been annihilated
by differentiating $g-1$ times with respect to $\epsilon'$ are most easily
computed
by expanding \sothree\ in powers of $\epsilon$:
$$\ZZ(\epsilon)={1\over 2(8\pi^2)^{g-1}}\sum_{k=0}^{g-2}
{{(-\pi^2\epsilon)}^k\over k!}(1-2^{3-2g+2k})\zeta(2g-2-2k)+O(\epsilon^{g-1}).
\eqn\occo$$

Using Euler's formula expressing $\zeta(2n)$ for positive integral
$n$ in terms of the Bernoulli number $B_{2n}$,
$$\zeta(2n)={(2\pi)^{2n}(-1)^{n+1}B_{2n}\over 2(2n)!}, \eqn\docco$$
\occo\ implies
$$\int_{{\cal M}'(-1)}\exp\left(\omega+\epsilon\Theta\right)
= (-1)^{g+1}\sum_{k=0}^{g-1}{\epsilon^k\over
k!}{(2^{2g-2-2k}-2)B_{2g-2-2k}\over
2^{3g-1}(2g-2-2k)!}.\eqn\sothreeform$$
(We have also used \coco\ and $B_0=1$ to get the term of order
$\epsilon^{g-1}$.)
This agrees with the final equation (29) of [\thaddeus] provided one notes
that the relation between Thaddeus's classes and ours is $\alpha=2\omega$,
$\beta=4\Theta$,
and that his ${\cal N}_g$ (which is the moduli space of flat $SU(2)$
connections on a once-punctured surface with monodromy $-1$ around the
puncture)
is an unramified $2^{2g}$-fold cover of our ${\cal M}'(-1)$ (which is the
moduli
space of flat $SO(3)$ connections on a non-trivial bundle).

In particular, \sothreeform\ is a polynomial in $\epsilon$ of degree $g-1$,
while on dimensional grounds it might have been of degree
${\rm dim}({\cal M})/4=(3/2)(g-1)$.  The fact that the higher coefficients
of the polynomial vanish is a reflection of the fact that $p_1({\cal M})^g
=0$  (a conjecture of Gieseker proved by F. Kirwan [\fkirwan]).

\sothreeform\ is not yet a complete answer for the intersection pairings
on ${\cal M}'(-1)$.  It is necessary also to include certain non-algebraic
cycles; we do this in \S4.5.

\subsection{Analogous Formulas For $SU(2)$}

Now we consider the case of gauge group $SU(2)$.
We start with
$${\partial^{g-1} Z(\Sigma,\epsilon)\over\partial\epsilon^{g-1}}
=  {(-1)^g\over (8\pi^2)^{g-1}}
\sum_{n=1}^\infty{\exp
\left(-{\epsilon n^2\over 4}\right)}={(-1)^g\over 2\cdot(8\pi^2)^{g-1}}
\left(-1+\sum_{n\in\IZ}\exp
\left(-{\epsilon n^2\over 4}\right)\right). \eqn\sutwo$$
Use of the Poisson summation formula now gives
$${\partial^{g-1}Z\over\partial\epsilon^{g-1}}
={(-1)^g\over 2\cdot(8\pi^2)^{g-1}}\cdot
\left(-1+\sqrt{4\pi\over \epsilon}\sum_{m\in\IZ}
\exp\left(-{(2\pi m)^2\over \epsilon}\right)\right). \eqn\otherform$$

The exponents of the exponentially small terms
in \otherform\ are in agreement with our expectations from \critval.
The novelty, compared to our discussion of $SO(3)$, is that the term on the
right hand side of
\otherform\ with $m=0$ does not vanish exponentially for small $\epsilon$.
As a result, the $(g-1)^{th}$ derivative of $Z$ is not a constant for
small $\epsilon$, but proportional to $\epsilon^{-1/2}$.
The general structure is thus
$$Z(\epsilon)=\sum_{k=0}^{g-2}a_k\epsilon^k+a_{g-3/2}\epsilon^{g-3/2}
+{\rm exponentially~small~terms}.\eqn\hchc$$
The coefficients in this expansion can easily be worked out as in
\occo.

As always, the terms in \hchc\
that do not vanish exponentially must be interpreted as the contribution
of $\mu^{-1}(0)$ to the localization formula.  The non-analyticity
of the contribution of $\mu^{-1}(0)$ reflects the fact that, for gauge
group $SU(2)$, $\mu^{-1}(0)$ is singular.  By studying the predictions
of the localization theory when $\mu^{-1}(0)$
is singular, it should be possible to interpret the exponent $g-3/2$ of
the singular term in terms of the singularities of the moduli space
${\cal M}$ of flat $SU(2)$ connections.

For non-trivial $SO(3)$ bundles, the contribution of $\mu^{-1}(0)$ to
$\ZZ(\epsilon)$ is a polynomial, whose coefficients are elementary multiples of
$\int_{\cal M'}\omega^{3g-3-2r}\Theta^r$.
In the $SU(2)$ case, such an interpretation cannot hold
as the contribution of $\mu^{-1}(0)$ is not a polynomial.
In some instances, Donaldson has shown that intersection pairings
(on singular four dimensional moduli spaces) analogous
to $\int\omega^{3g-3-2r}\Theta^r$
are well-defined only for small enough $r$.  Our considerations
here perhaps give a new framework for this phenomenon: the function
$Z(\epsilon)$ is defined in any case, but the extent to which it has an
asymptotic expansion in integral powers of $\epsilon$ and the interpretation
of the coefficients as intersection pairings depend on details of the classical
geometry of the moduli space.

\section{Genus Zero}

Now we are going to look more closely at the behavior of the partition function
for the case that $\Sigma$ is a closed Riemann surface of genus zero.
We consider arbitrary compact, connected (but not necessarily simply
connected) gauge group $H'$.   The general formula for the partition
function, specialized to genus zero, is
$$ \ZZ(\Sigma,\epsilon;u)=
{1 \over \# \pi_1(H')}\cdot
\left({(2\pi)^{\rm \dim H'}\over{\rm Vol}(H')}\right)^{2}
\cdot \sum_\al(\dim\alpha)^2\lambda_\al(u^{-1})
\cdot\exp\left(-\epsilon\left({C_2(\al)\over 2}+t\right)\right).
\eqn\xox$$

If one takes $H'$ to be $SU(2)$ or $SO(3)$, then the right hand side
of \xox\ is essentially the derivative of a theta function, rather than
the $g-1$-fold integral of a theta function
considered earlier.
\xox\ can be expanded as a sum of contributions of critical points,
using similar arguments to those we gave above for $g\geq 1$; this
is left to the reader.  Our intention here
is to analyze closely the contribution of $\mu^{-1}(0)$. We will do this
without restriction on $H'$.  We actually will only look at the leading
behavior for $\epsilon\to 0$, so
we can set $t=0$.  (This is fortunate as we have not yet
determined $t$ for general $H'$.)
The representation of $\ZZ$ that we will use appears in the work of
Fine [\fine] and Forman [\forman].

Actually, $\mu^{-1}(0)$ is empty in genus zero unless $u=1$,
since a flat connection on a two sphere with one point
deleted cannot have a non-trivial monodromy around the puncture.
So \xox\ should vanish
exponentially except for $u=1$.  We will verify this presently.

\def\ff{{\rm Fun}(H)}
Let $\ff$ be the space of functions on the $H$ manifold, regarded
as an $H_L\times H_R$ module (with $H_L\times H_R$ being two copies of
$H$, acting on $H$ by $h\to ahb^{-1}$).
The decomposition of $\ff$ in irreducible $H_L\times H_R$
modules is (by the Peter-Weyl theorem)
$$\ff\cong \oplus_\alpha \alpha\otimes\overline\alpha, \eqn\nsns$$
where the sum runs over isomorphism classes of irreducible $H_L$ modules
$\alpha$, and $\overline\alpha$ is the  complex conjugate $H_R$ module.

We recall that we defined the quadratic Casimir operator $C_2$ by
$C_2=-\sum_aT_a^2$, where $T_a$ runs over a basis of ${\cal H}$ orthonormal
with respect to a certain invariant metric.  The same metric determines
a Laplace operator $\Delta $ on $H$.
Since $\alpha$ and $\overline\alpha$ have the same value of
$C_2$, it follows from \nsns\ that
the quadratic Casimirs of $H_L$ and
$H_R$ coincide as operators on $\ff$; and moreover, both are equal to $\Delta$.

If, therefore, $u$ is in the center of $H$ and
$T(u^{-1})$ is the operator of left multiplication by
$u^{-1}\in H$, then using the Peter-Weyl theorem
$$\Tr_{\ff}T(u^{-1})\exp\left({-\epsilon\Delta\over 2}\right) =\sum_\alpha
(\dim\alpha)^2\lambda_\alpha(u^{-1})\exp\left(-{\epsilon C_2(\al)\over
2}\right)
   .
\eqn\nxnx$$   We have used the fact that $\Delta$ acts on the representation
$\alpha$ as $C_2(\alpha)$.
The right hand side is, up to an elementary constant, the desired
function $\ZZ(\Sigma,\epsilon;u)$.

We can now see the expected exponential vanishing of $\ZZ$ for $\epsilon \to 0$
with $u\not= 1$.
The left hand side of \nxnx\ is essentially
a matrix element of the heat kernel $\exp(-\epsilon\Delta/2)$.  By the
general theory of the short time behavior of the heat kernel,
it vanishes exponentially for $\epsilon\to 0$
and any fixed
$u\not= 1$, with an exponent determined by the length of the shortest
geodesic on $H$ from $1$ to $u$.

It remains to consider the case $u=1$.
According
to the general theory of the heat kernel, the
left hand side of \nxnx\ is for $u=1$
asymptotic for small $\epsilon$ to
${\rm Vol}(H)/(2\pi\epsilon)^{\dim(H)/2}$.
Using this in \xox\ (and recalling that ${\rm Vol}{(H)}={\rm Vol}(H')
\cdot \#\pi_1(H')$), we get the small $\epsilon$ asymptotic behavior
$$ \ZZ(\Sigma,\epsilon;u)\sim {1\over {\rm Vol}(H')}\cdot\left({(2\pi)^3\over
\epsilon}\right)^{\dim H'/2}. \eqn\noxox$$

\subsection{Comparison To The Localization Formula}

Now we compare to the localization formula.  As always, $\mu^{-1}(0)$
consists of flat connections.  In genus zero, every flat connection
is gauge equivalent to the trivial connection.
The group $G'$ of gauge transformations does
not act freely on the space of flat connections;
the trivial connection, for instance, is invariant under
the finite dimensional group $H'$ of constant gauge transformations,
and every flat connection has a stabilizer isomorphic to this.
So $\mu^{-1}(0)$ is a copy of $G'/H'$.  A neighborhood of
$\mu^{-1}(0)$ in the space ${\cal A}$ of connections can therefore
be modeled on $T^*(G'/H')$.  (By using the relation of connections
and complex structures, one can even -- after picking a complex structure
on $\Sigma$  -- identify a dense
open set in ${\cal A}$ with $T^*(G'/H')$.)  Therefore, the problem
of identifying the contribution of $\mu^{-1}(0)$ to $\ZZ$ is an infinite
dimensional version of the problem
that was solved in finite dimensions
in equation \dnd.

Two-dimensional Yang-Mills theory is equivariant
integration over ${\cal A}$ of
$\exp(\overline \omega)$, where $\overline\omega=\omega
+i\sum_a\phi^a\mu_a$ is the equivariant extension of
the symplectic form $\omega$ introduced in equation \urgo.
Thus the form $\alpha$ of equation
\dnd\ can be identified with $\exp(\overline\omega)$.
Since $\omega$ restricts to zero on $\mu^{-1}(0)$, and the same is of
course also true of $\mu$, the reduced form $\alpha'$ of
equation \dnd\ is 1.
Therefore \dnd\ identifies the contribution of $\mu^{-1}(0)$ to the
functional integral as
$${1\over {\rm Vol}(H')}\int {\d\phi_1\dots \d\phi_{\dim H'}
\over (2\pi)^{\dim H'}}\exp\left({\epsilon\over 8\pi^2}\Tr  \phi
^2\right). \eqn\nnsns$$
(The factor of $8\pi^2$
is inherited from the definition of the Yang-Mills
Lagrangian in \physlag.)
Upon performing explicitly
the Gaussian integral in \nnsns,
we recover the asymptotic expression
of \noxox, as desired.

\section{Pairings Of Non-Algebraic Cycles For $SO(3)$}

The cohomology of the smooth $SO(3)$ moduli space ${\cal M}'(-1)$ is
known [\newstead,\abott] to be generated by the classes
$\omega$ and $\Theta$, whose intersection pairings have been determined
in equation \sothreeform\ above, along with certain non-algebraic cycles,
which we will now incorporate.

The basic formula that we will use is equation \kuko\ from \S3.2:
$$\eqalign{
\langle\exp\left(\omega+\epsilon\Theta\right)\cdot \beta \rangle'
={1\over {\rm vol}(G)}
\int DA\,\,D\psi\,\,D\phi \,\,\exp&\left( {1\over 4\pi^2}
\int_\Sigma\Tr\left(i\phi F+{1\over 2}\psi\wedge\psi\right)
\right.\cr&~~~~~~~\left.+{\epsilon\over 8\pi^2}\int_\Sigma
\d\mu\Tr\phi^2\right)
\cdot \beta.\cr} \eqn\kkuko$$
We recall that $\langle ~~~\rangle'$ coincides with integration over
moduli space, up to terms that vanish exponentially for $\epsilon\to 0$.

Note that $\psi$ is a free field, with a Gaussian measure, and the
``trivial'' propagator
$$\langle\psi_i^a(x)\psi_j^b(y)\rangle =-4\pi^2
\epsilon_{ij}\delta^{ab}\delta^2(x-y).
     \eqn\kukuko$$
This will make life easy.

The new cycles that we must incorporate have the following description,
from \S3.3.  For every circle $C\subset \Sigma$ there is a quantum field
operator
$$ V_C={1\over 4\pi^2}\int_C\Tr\phi\psi.        \eqn\mcmcll$$
It represents a three dimensional class on moduli space; this
class depends only on the homology class of $C$.
As the algebraic cycles are  even dimensional, non-zero intersection
pairings are possible only with an even number of the $V_C$'s.  The
first case is $\langle \exp(\omega+\epsilon\Theta)\cdot V_{C_1}V_{C_2}\rangle'
$, with two oriented circles $C_1,$ $C_2$ that we can suppose to intersect
transversely in finitely many points.  So we consider
$$\eqalign{
\langle\exp\left(\omega+\epsilon\Theta\right)\cdot V_{C_1}V_{C_2}\rangle'
&={1\over {\rm vol}(G)}
\int DA\,\,D\psi\,\,D\phi \,\,\exp\left( {1\over 4\pi^2}
\int_\Sigma \Tr\left(i\phi F+{1\over 2}\psi\wedge\psi\right)
\right.\cr&~~~~~~~\left.+{\epsilon\over 8\pi^2}\int_\Sigma
\d\mu\Tr\phi^2\right)
\cdot {1\over 4\pi^2}\int_{C_1}\Tr\phi\psi\cdot{1\over 4\pi^2}
\int_{C_2}\Tr\phi\psi.\cr} \eqn\kkukko$$
Upon performing the $\psi$ integral, using \kukuko, we see that
this is equivalent to
$$\eqalign{{1\over {\rm vol}(G)}
\int DA\,\,D\psi\,\,D\phi \,\,\exp &\left( {1\over 4\pi^2}
\int_\Sigma \Tr\left(i\phi F+{1\over 2}\psi\wedge\psi\right)
\right.\cr &~~~~~~~\left.+{\epsilon\over 8\pi^2}\int_\Sigma
\d\mu\Tr\phi^2\right)
\cdot \sum_{P\in C_1\cap C_2}{-\sigma(P)\over 4\pi^2}\Tr\phi^2(P)
.\cr} \eqn\kkukko$$
Here $P$ runs over all intersection points of $C_1$ and $C_2$, and
$\sigma(P)=\pm 1$ is the oriented intersection number of $C_1$ and $C_2$ at
$P$.  Since the cohomology class of $\Tr\phi^2(P)$ is independent of
$P$, and equal to that of $\int_\Sigma \d\mu \Tr\phi^2$, \kkukko\ implies
$$\eqalign{
\langle\exp\left(\omega+\epsilon\Theta\right)\cdot V_{C_1}V_{C_2}\rangle'
={1\over {\rm vol}(G)}
&\int DA\,\,D\psi\,\,D\phi \,\,\exp \left( {1\over 4\pi^2}
\int_\Sigma \Tr\left(i\phi F+{1\over 2}\psi\wedge\psi\right)
\right.\cr  & \left.+{\epsilon\over 8\pi^2}
\int_\Sigma\Tr\phi^2\right)\cdot\left(-{\#(C_1\cap C_2)\over 4\pi^2}
\int_\Sigma\Tr\phi^2\right),
\cr}\eqn\okkukko$$
with $\#(C_1\cap C_2)=\sum_P\sigma(P)$
the algebraic intersection number of $C_1$ and
$C_2$.
\okkukko\ is equivalent to
$$\langle \exp(\omega+\epsilon\Theta)V_{C_1}V_{C_2}\rangle'
=-2\#(C_1\cap C_2)\cdot
{\partial\over\partial \epsilon}\langle\exp(\omega+\epsilon\Theta)\rangle',
\eqn\mmdmd$$
which interpreted in terms of intersection numbers gives in particular
$$\int_{{\cal M}'(-1)}\exp(\omega+\epsilon\Theta)V_{C_1}V_{C_2}
=-2\#(C_1\cap C_2){\partial\over\partial
\epsilon}\int_{{\cal M}'(-1)}\exp(\omega+\epsilon\Theta).  \eqn\dmmdm$$
Of course, the right hand side is known from \sothreeform.

The generalization to an arbitrary number of $V$'s is almost immediate.
Consider oriented circles $C_\sigma,$ $\sigma
=1\dots 2g$, representing a basis of $H_1(\Sigma,\IZ)$.  Let
$\gamma_{\sigma\tau}
=\#(C_\sigma\cap C_\tau)$ be the matrix of intersection numbers.
Introduce anticommuting parameters
$\eta_\sigma$, $\sigma=1\dots 2n$.  I claim
$$\int_{{\cal M}'(-1)}\exp\left(\omega+\epsilon\Theta+\sum_{\sigma=1}^{2g}
\eta_\sigma V_{C_\sigma}\right) =\int_{{\cal M}'(-1)}\exp\left(\omega+
\widehat \epsilon\Theta\right),
\eqn\bcbcbc$$
with
$$\widehat\epsilon=\epsilon-2\sum_{\sigma<\tau}\eta_\sigma\eta_\tau
\gamma_{\sigma\tau}.
\eqn\bhbh$$
The computation leading to this formula is a minor variant of the one
we have just done.  The left hand side of \bcbcbc\ is equal (up to terms
that vanish exponentially for $\epsilon\to 0$) to
$$\eqalign{
{1\over {\rm vol}(G)}
\int DA\,\,D\psi\,\,D\phi \,\,\exp&\left( {1\over 4\pi^2}
\int_\Sigma\Tr\left(i\phi F+{1\over 2}\psi\wedge\psi\right)
\right.\cr&~~~~~~~\left.+{\epsilon\over 8\pi^2}\int_\Sigma
\d\mu\Tr\phi^2+{1\over 4\pi^2}\sum_{\sigma=1}^{2n}\eta_\sigma\int_{C_\sigma}
\Tr\phi\psi
\right). \cr}\eqn\hhurko$$
Shifting $\psi$ to complete the square, and then performing the Gaussian
integral over $\psi$, this becomes
$$\eqalign{
{1\over {\rm vol}(G)}
\int DA\,\,D\psi\,\,D\phi \,\,\exp&\left( {1\over 4\pi^2}
\int_\Sigma\Tr\left(i\phi F+{1\over 2}\psi\wedge\psi\right)
\right.\cr&~~~~~~~\left.+{\widehat\epsilon
\over 8\pi^2}\int_\Sigma
\d\mu\Tr\phi^2
\right). \cr}\eqn\hhhurko$$
The polynomial part of this is the right hand side of \bcbcbc.

The method of eliminating the non-algebraic cycles that we have
just described is somewhat analogous to Proposition (26) of
[\thaddeus].
Our formulas \dmmdm\ and \bcbcbc\ are equivalent to the formula
given by Thaddeus in Proposition (26), modulo the evaluation \sothreeform\
of intersection numbers of algebraic cycles
and the fact that our $V_C$ is the same as Thaddeus's $\psi_C$.

\section{Casimir Operators For Arbitrary Lie Groups}

Let $\widetilde\phi=\phi/4\pi^2$, and let $Q(\widetilde\phi)$ be
an invariant polynomial
in $\phi$, homogeneous of degree $t$.  $Q$ should correspond to an operator
$\widehat Q$ in two-dimensional quantum Yang-Mills theory, and in
\ibo, we showed that, with a particular definition of the quantum theory,
this is just the Casimir operator $Q(-iT)$.

In general, the passage from classical to quantum mechanics is uniquely
determined only up to a renormalization of the various operators and
parameters.
In a theory which is as strongly ultraviolet convergent as two-dimensional
Yang-Mills theory, the only ambiguity is ``normal ordering''; different
ways of defining the quantum operator corresponding to $Q(\widetilde\phi)$
will differ only by terms that can be considered to come from the addition
to $Q$ of invariant polynomials of lower degree.  Thus, in general,
with an arbitrary method of defining the theory,
$\widehat Q=Q(-iT)\,+\,$lower order Casimir operators.

In principle, the topological regularization of \S3 should uniquely
determine the normal-ordering recipe; I will leave this as an
interesting open problem.
But part of the story is easy to discern.
The topological regularization certainly preserves equation \firsteqn,
so we should restrict ourselves to normal-ordering prescriptions compatible
with this.\foot{The extension of the formalism
to incorporate \secondeqn\ and \thirdeqn\ will be explained in \S5.}
This means simply that $Q\to\widehat Q$ must be a ring homomorphism
from invariant polynomials on the Lie algebra to quantum operators.
Thus, it is sufficient to determine $\widehat Q$ with $Q$ ranging over
a set of generators of the ring of invariant polynomials.

For instance, for $SU(2)$, this ring is a polynomial ring with one
generator,
which we can take to be
$Q(\widetilde\phi)=\Tr\widetilde\phi^2$. In this case,
\ibo\ gives $\widehat Q=-\sum_a\Tr T_a^2=C_2$,
while a lower order Casimir operator would have to be a constant.
This one constant is the only normal-ordering ambiguity for $SU(2)$.
It was called $t$ in equation \nuuf.  Ideally, $t$ should be determined
by an {\it a priori} calculation using a regularization in which
the relation of \S3 between the physical and topological theories is valid.
We did so more pragmatically at the beginning of \S4.3
to ensure a particular consequence of this relation (certain functions
should be polynomials).

In general, for a compact
Lie group $H$ of rank $r$, the ring of invariant polynomials
is a polynomial ring in $r$ generators
(see [\zel,\S126]).  For any given $H$, there are therefore
finitely many analogs of the normal ordering constant $t$.
We will now state a generalization of the prescription $t=1/4$
to arbitrary Lie groups and arbitrary Casimirs.

First we give a convenient restatement of the situation for $SU(2)$.
Let us call $v$ the highest weight of the two dimensional representation
of $SU(2)$.  The highest weight of the $n$ dimensional representation
$\alpha_n$ is then $h=(n-1)v$.  One half the sum of the positive roots of
$SU(2)$ is $\delta = v$, so $h+\delta = n v$.
The generator
$Q=\Tr\widetilde\phi^2$ of the ring of invariants corresponds, if $t=1/4$,
to the quantum operator  $\widehat Q(\alpha_n)
=n^2/2$.  Since the Weyl group of $SU(2)$ is $\IZ/2\IZ$, acting
by $v\to -v$, this can be described as follows: for a representation of
highest weight $h$, $\widehat Q$ is a homogeneous, Weyl-invariant polynomial
in $h+\delta$, which agrees with the usual quadratic Casimir
$Q(-iT)$ up to terms of lower order.

This formulation can be immediately generalized.   Let $Q(\widetilde\phi)$
be any homogeneous invariant polynomial of degree $n$
on the Lie algebra of a compact
simple Lie group $H$.  According to [\zel,\S126, Theorem 7],
the corresponding Casimir $Q(-iT)$ is equal,
on an irreducible  representation $\alpha_h$ of highest weight $h$,
to $Q'(h+\delta)$, where $Q'$ is a Weyl-invariant
polynomial of degree $n$.
Moreover $Q'(h+\delta)=Q(h+\delta)+\dots$, where the ``$\dots$'' are
lower order Weyl-invariant polynomials, which coincide, by essentially
the same theorem, with some Casimirs of lower order.  Since the ability
to add such terms is precisely the normal ordering ambiguity, we can
pick a normal ordering recipe in which, acting on $\alpha_h$,
$$\widehat Q = Q(h+\delta).         \eqn\normalorderingrecipe$$
For the time being, this is an arbitrary ansatz.  However, this
choice will enter at the end of \S5 in verifying
consequences of the relation of \S3 between
the physical and topological theories.

\subsection{Generalization Of The Path Integral}

We now want to evaluate the following generalization of the conventional
Yang-Mills path integral \nxnxnx:
$$\int DA \,\,D\phi \,\,\,\exp\left({i\over 4\pi^2}\int_\Sigma
\Tr\phi F +\int_\Sigma Q(\wph)\right), \eqn\murgo$$
with $Q(\wph)$ an arbitrary invariant polynomial on ${\cal H}$
(with some positivity properties to ensure convergence of the integral,
or with the higher than quadratic terms in $Q$ having nilpotent coefficients
to avoid such questions).  The path integral can be evaluated by summing
over the same physical states as before.  The only novelty is that
the Hamiltonian is different: it is now $H=-\widehat Q$. With our
normal-ordering recipe, the generalization
of \ooxox\ is then
$$\widetilde Z(\Sigma,Q;u)=
{1 \over \# \pi_1(H')}\cdot
\left({{\rm Vol}(H')\over (2\pi)^{\rm dim H'}}\right)^{2g-2}
\cdot \sum_h{\lambda_h(u^{-1})\cdot\exp\left(Q(h+\delta)\right)
\over d(h)^{2g-2}}. \eqn\deffor$$
Here $h$ runs over dominant weights of $H$ (which are in one-to-one
correspondence, of course, with isomorphism classes of irreducible
$H$ modules $\alpha(h)$), $d(h)$ is the dimension of $\alpha(h)$,
and $\lambda_h = \lambda_{\alpha(h)}$.

\chapter{The Intersection Ring Of The Moduli Space}

In \S4, we computed the intersection pairings on the moduli space
of flat $SO(3)$ connections on a non-trivial bundle over a surface
of genus $\geq 2$.  We now want to extend this computation to other groups.
We consider an arbitrary compact connected gauge group $H'$, with simply
connected cover $H$.  We work on an arbitrary $H'$ bundle $E'(u)$ over
a  closed oriented surface $\Sigma$ of genus $g$.  We will evaluate
the quantum field theory partition function in general, and then interpret
it in terms of intersection pairings for the smooth cases.

The basic formula that we will use is our friend
\kuko:
$$\eqalign{
\langle\exp\left(\omega+\epsilon\Theta\right)\cdot \beta \rangle'
={1\over {\rm vol}(G')}
\int DA\,\,D\psi\,\,D\phi \,\,\exp&\left( {1\over 4\pi^2}
\int_\Sigma\Tr\left(i\phi F+{1\over 2}\psi\wedge\psi\right)
\right.\cr&~~~~~~~\left.+{\epsilon\over 8\pi^2}\int_\Sigma
\d\mu\Tr\phi^2\right)
\cdot \beta.\cr} \eqn\kukox$$
In \kukox, $\beta$ is supposed to be an equivariant differential form
with a polynomial dependence on $\phi$.

Since exponentials will be much
more convenient than polynomials, we will
resort to the following device.  We introduce bosonic and fermionic
variables $\delta_i$
which are nilpotent with $\delta_i^{n_i}=0$ for some unspecified $n_i$ if
$\delta_i$ is bosonic
(or $n_i=2$ if $\delta_i$ is fermionic).
We call such variables formal variables.
We  write $\beta=\exp(\sum_i\delta_i
\beta_i)$, where the $\beta_i$ have only a polynomial dependence on $\phi$
and the $\delta_i$ are formal variables.  Obviously, it is sufficient to
study \kukox\ for such $\beta$'s.  Taking the limit of $n_i\to\infty$
(for all $i$ such that $\delta_i$ is bosonic),
$\beta$ becomes a formal power series.

The localization principle can be applied to \kukox\ with such $\beta$,
since, for any $n_i$, $\beta$ is polynomial in
$\phi$.  After performing the integral in \kukox, we get a function
$g(\delta_i)$ of the formal variables which is, of course, a polynomial
for any given $n_i$ and becomes a formal power series for $n_i\to\infty$.
Under certain conditions, the localization theorem implies (and we will
verify later) that these formal power series are really polynomials.

Let us recall from \S3.3 what are the possible $\beta_i$.  For any
degree $r$ invariant
polynomial $Q$ on ${\cal H}$, we have equivariant differential forms
$$Q_{(0)}=\int_\Sigma \d\mu\,\,\,Q(\phi) \eqn\firone$$
and
$$Q_{(2)}=\int_\Sigma\left(i{\partial Q\over\partial\phi^a}F^a+{1\over 2}
{\partial^2 Q\over\partial\phi^a\partial\phi^b}\psi^a\wedge\psi^b\right)
             \eqn\girone$$
of degree $2r$ and $2r-2$.
Also, for any circle $C\subset\Sigma$, we have the form of degree $2r-1$
$$Q_{(1)}(C)=-\int_C{\partial Q\over\partial\phi_a}\psi^a. \eqn\tirone$$

Now, let $Q(\phi)$ be an invariant polynomial of the form
$$Q(\phi)={1\over 8\pi^2}\Tr\phi^2+\sum_i\delta_iq_i(\phi), \eqn\irone$$
where the $\delta_i$ are formal variables, and the $q_i(\phi)$ are homogeneous
of degree  $\geq  3$.

Let $T(\phi)$ be an invariant polynomial of the form
$$T(\phi)={\epsilon\over 8\pi^2}\Tr\phi^2+\sum_i\delta'_it_i(\phi),
   \eqn\porky$$
with the $\delta'_i$ formal variables and $t_i(\phi)$ homogeneous
of degree $\geq 3$.

Let $C_\rho\subset \Sigma$, $\rho=1\dots 2g$ be oriented circles generating
$H_1(\Sigma,\IZ)$, with intersection pairings $\gamma_{\sigma\tau}
=\#(C_\sigma\cap C_\tau)$.  For each $\rho$, pick an invariant
polynomial
$$S^\rho(\phi)=\sum_i\eta_i^\rho s^\rho_i, \eqn\nork$$
where the $s^\rho_i$ are invariant homogeneous polynomials and
the $\eta_i^\rho$ are anticommuting (and so in particular nilpotent)
parameters.

We aim to compute
$$\left\langle\exp\left(Q_{(2)}+T_{(0)}+\sum_\rho
S^\rho_{(1)}(C_\rho)\right)
\right\rangle'.               \eqn\norker$$

It is convenient to introduce
$$\widehat\phi_a=4\pi^2{\partial Q\over\partial \phi^a}. \eqn\orker$$
We will first evaluate \norker\ under the restriction
$$\det\left({\partial \widehat\phi^a\over\partial\phi^b}\right) = 1.
              \eqn\yorker$$
Modulo the elementary identities noted at the end of \S3, there
actually is no loss of information
in evaluating \norker\ only for $Q$ such that \yorker\ holds.
However, in any event,
after carrying out the calculation assuming \yorker, we will then
relax this requirement and consider the general case.

\subsection{The Computation}

The basic formula \kukox\ equates \norker\ with the following path integral:
$$\eqalign{{1\over {\rm vol}(G')}\int DA\,\,\,
D\psi\,\,\,D\phi\,\,\,&\exp\left(\int_\Sigma\left(i{\partial Q\over\partial
\phi^a}F^a+{1\over 2}{\partial^2Q\over\partial\phi^a\partial\phi^b}\psi^a\wedge
\psi^b\right)\right.\cr &-\left.
\sum_\sigma \oint_{C_\sigma}{\partial S^\sigma
\over\partial\phi^a}\psi^a+\int_\Sigma \d\mu
\,\,T(\phi)\right).\cr}\eqn\omero$$

\def\wrt{\widehat T}
First we carry out the integral over $\psi$.  Because of
\yorker, the $\psi$ determinant coincides with what it would be if
$Q=\Tr\phi^2/8\pi^2$.  As we have discussed in connection with \msms,
this determinant just produces the standard symplectic measure on the
space ${\cal A}$ of connections; this measure we conventionally
(but somewhat inconsistently) call $DA$.  Let $(\partial^2Q)^{-1}$ be
the inverse matrix to the matrix $\partial^2Q/\partial\phi^a\partial\phi^b$,
and let
$$\wrt(\phi) = T(\phi)-\sum_{\sigma<\tau}\gamma_{\sigma\tau}{\partial S^\sigma
\over\partial\phi^a}{\partial S^\tau\over\partial\phi^b}(\partial^2Q)^{-1}
_{ab}. \eqn\umero$$
The second term arises, as in the derivation of \okkukko, in shifting
$\psi$ to complete the square in
\omero.
Then integrating out $\psi$ gives
$${1\over {\rm Vol}(G')}\int DA\,\,\,D\phi\,\,\,
\exp\left(i\int_\Sigma{\partial Q\over \partial \phi^a}F^a
+\int_\Sigma \d\mu \,\,\,\,\wrt(\phi)\right). \eqn\ymero$$
Now change variables from $\phi$ to $\widehat\phi$, defined in \orker.
The Jacobian for this change of variables is 1 because of \yorker.
Because the $\delta_i$ are nilpotent, the transformation is invertible;
the inverse is given by some functions $\phi^a=W^a(\widehat \phi)$.
After the change of variables, \ymero\ becomes
$${1\over {\rm Vol}(G')}\int DA\,\,\,D\widehat\phi\,\,\,
\exp\left({i\over 4\pi^2}\int_\Sigma\Tr\widehat \phi F
+\int_\Sigma \d\mu \,\,\,\,\wrt\comp W(\widehat\phi)\right).\eqn\zymero$$
This is a path integral
of the type that we evaluated in equation \deffor.

We observed in \S4 that in canonical quantization, $\widehat\phi^a/4\pi^2$ is
identified with the group generator $-iT^a$.
To avoid repeated factors of $4\pi^2$, define an invariant function
$V$ by
$W(\widehat \phi)  =  V(\widehat \phi/4\pi^2)$.
The invariant
function $\wrt\comp W(\widehat\phi)$ corresponds
in the quantum theory (using the normal-ordering prescription of
\S4.6) to the operator that on a representation of highest
weight $h$ is equal to $\wrt\comp V(h+\delta)$, with $\delta$ equal
to half the sum of the positive roots.  Borrowing the result of \deffor,
the explicit evaluation of \zymero\ gives
$${1\over \#\pi_1(H')}\left({\rm Vol}(H')\over (2\pi)^{\dim(H')}\right)^{2g-2}
\sum_h{\lambda_h(u^{-1})\exp
\left({\wrt\comp V}(h+\delta)\right)
\over d(h)^{2g-2}},\eqn\pimero$$
with $h$ running over dominant weights and $\delta$ as above.

Now we want to relax the assumption of \yorker.  We will do so somewhat
informally, but the result could be justified using
the regularization of \S3.

\yorker\ was used twice. The determinant in the $\psi$ integral would
be formally, if \yorker\ is not assumed,
$$\prod_{x\in\Sigma}\det\left({\partial^2Q'\over\partial\phi^a\partial\phi^b}
\right)         ,\eqn\timero$$
times the determinant for $Q=\Tr\phi^2/8\pi^2$.
We have set $Q'=4\pi^2Q$.
The factors in \timero\ are all equal up to coboundaries (since more generally,
for any invariant function $U$ on ${\cal H}$, $U(\phi(P))$ is cohomologous
to $U(\phi(P'))$, for $P,P'\in\Sigma$, according to equation \umitsu).
Of course, this infinite product of essentially equal factors
diverges unless \yorker\ is assumed.
The Jacobian in the changes of variables from $\phi$ to $\widehat\phi$
is formally
$$\prod_{x\in\Sigma}\left(\det\left({\partial^2Q'\over\partial\phi^a\partial
\phi^b}\right)\right)^{-1}.             \eqn\imero$$
Formally, these two factors appear to cancel, but this cancellation should
be taken to mean only that the result is finite, not that it equals one.
The number of factors in \timero\ should be interpreted as $N_1/2$, half the
dimension of the space of one-forms.  The number of factors in \imero\ should
be interpreted as $N_0$, the dimension of the space of zero-forms.
The difference $N_1/2-N_0$ is $-1/2$ the Euler characteristic of $\Sigma$,
or $g-1$.  Thus the product of \timero\ and \imero\ should be interpreted
as $\det\left(\partial^2Q'/\partial\phi^a\partial\phi^b\right)^{g-1}$.
A convenient function cohomologous to this is $\exp\left(\int_\Sigma
(g-1)\ln\det\left(\partial^2Q'/\partial\phi^a\partial\phi^b\right)\right)$.
\foot{In a calculation using the regularization of \S3, this factor
would arise from the one loop chiral anomaly.  It would appear in the form
of a factor $\exp\left(-\int_\Sigma \d\mu \,\,(R/8\pi)\cdot
\ln\det(\partial^2Q'/\partial\phi^a
\partial\phi^b)\right)$, with $R$ the scalar curvature of the metric that
enters in the cohomological Lagrangians of \S3.}

The sole result of relaxing \yorker\
is accordingly that \zymero\ becomes
$${1\over {\rm Vol}(G')}\int DA\,\,\,D\widehat\phi\,\,\,
\exp\left({i\over 4\pi^2}\int_\Sigma\Tr\widehat \phi F
+\int_\Sigma \d\mu \,\,\,\widetilde T\comp
W(\widehat\phi)\right),\eqn\zzymero$$
with
$$\widetilde T=\wrt+(g-1)\ln\det\left({\partial^2Q'\over\partial\phi^a\partial
\phi^b}\right)=T
-\sum_{\sigma<\tau}\gamma_{\sigma\tau}{\partial S
^\sigma\over\partial\phi^a}{\partial S^\tau
\over\partial\phi^b}(\partial^2Q)^{-1}
_{ab}+(g-1)\ln\det\left({\partial^2Q'\over\partial\phi^a\partial\phi^b}\right).
          \eqn\zzzymero$$
The evaluation of the path integral therefore leaves in general
not quite \pimero\ but
$${1\over \#\pi_1(H')}\left({\rm Vol}(H')\over (2\pi)^{\dim(H')}\right)^{2g-2}
\sum_h{\lambda_h(u^{-1})\exp
\left({\widetilde T\comp V}(h+\delta)\right)
\over d(h)^{2g-2}}.\eqn\pppimero$$

\subsection{The Topological Conclusion}

For $\epsilon>0$ and the other parameters nilpotent, the path integral
can be evaluated as we have just done regardless of possible singularities
of the moduli space ${\cal M}'(u)$ of flat connections on $E'(u)$.
If $H'$, $u$, and $\Sigma$ are such that the space of flat connections
is smooth, and the gauge group acts freely on it (at least modulo its center),
then according to
the theory of \S2,
\pppimero\ is a polynomial in $\epsilon$ and the formal variables,
modulo terms that vanish exponentially for $\epsilon\to 0$.  Moreover, this
polynomial then
has an interpretation in terms of intersection numbers on moduli
space:
$$\eqalign{
\int_{{\cal M}'(u)}&\exp\left(Q_{(2)}+\sum_\sigma S^\sigma_{(1)}
(C_\sigma)+T_{(0)}\right)\cr &=  {\# Z(H')
\over \#\pi_1(H')}\left({\rm Vol}(H')\over (2\pi)^{\dim(H')}\right)^{2g-2}
\sum_h{\lambda_h(u^{-1})\exp\left(\widetilde T\circ V(h+\delta)\right)
\over d(h)^{2g-2}}+\dots,\cr}\eqn\jimero$$
where ``$\dots$'' are exponentially small terms.
($\# Z(H')$, the order of the isomorphism group of a generic connection,
is the usual factor relating the path integral to the intersection theory.)
This formula generalizes the computations we made in \S4
for the case of $H=SU(2)$, $H'=SO(3)$, $u=-1$.  It applies notably for
$H=SU(N)$,
with center $\Gamma\cong \IZ/N\IZ$,
$H'=SU(N)/\Gamma$, and $u$ a generator of $\Gamma$.

Even when $H',u$, and $\Sigma$ are such that the space of flat connections
is not smooth (or the gauge group does not act freely on it), certain
intersection pairings on ${\cal M}'(u)$ are well-defined topologically.
A more careful study of the contribution of $\mu^{-1}(0)$ to the localization
formula will probably show that these are all given by the
asymptotic expansion of \jimero.

An important check is that \jimero\ is compatible with the general
topological relations of equations \firsteqn--\thirdeqn.
To write out these relations, let us abbreviate
$$Z(T,S^\sigma,Q)=\int_{{\cal M}'(u)}\exp\left(Q_{(2)}+
\sum_\sigma S^{\sigma}_{(1)}(C_\sigma)+T_{(0)}\right). \eqn\ormox$$
Then \firsteqn\ amounts to the statement that for any invariant polynomials
$A,B$
$$\left.{\partial \over\partial\epsilon}Z(T+\epsilon AB,S^\sigma,Q)
\right|_{\epsilon=0}
=\left.{\partial^2
\over\partial\alpha\partial\beta}Z(T+\alpha A+\beta B,S^\sigma,Q)
\right|_{\alpha=\beta=0}.  \eqn\junox$$
This is easily verified.  \secondeqn\ amounts to
the statement that for any $A$ and $B$ and any $\tau$
$$\left.{\partial\over\partial\epsilon}Z(T,S^\sigma+\delta^\sigma_\tau
\epsilon AB,Q)\right|_{\epsilon=0}
=\left.\left({\partial^2
\over\partial\alpha\partial\beta}Z(T+\alpha A,S^\sigma+\delta^\sigma_\tau\beta
 B,Q) +A\leftrightarrow B\right)\right|_{\alpha=\beta=0}.
\eqn\unox$$
Here $\alpha$ is commuting and $\beta,\epsilon$ are anticommuting.
This is also easy to verify.
The last relation \thirdeqn\ amounts to
$$\eqalign{\left.{\partial\over\partial\epsilon}
Z(T,S^\sigma,Q+\epsilon AB)\right|_{\epsilon=0}
=& \left.\left({\partial^2\over
\partial\alpha\partial\beta} Z(T+\alpha A,S^\sigma,Q+\beta B)+
A\leftrightarrow B  \right)\right|_{\alpha=\beta=0}
\cr &
+\sum_{\tau,\nu =1}^{2g}\gamma_{\tau\nu}{\partial^2\over\partial
\alpha'\partial\beta'}
Z(T,S^\sigma+\delta^\sigma_\tau\alpha' A+\delta^\sigma_\nu\beta' B,Q)
|_{\alpha'=\beta'=0}.\cr}\eqn\xuxo$$
(Here $\alpha'$ and $\beta'$ are anticommuting.)
The verification is straightforward but a little longer.  It is only
here that the ``anomaly'' term -- involving the determinant of
$\partial^2Q'/\partial\phi^2$  -- plays a role.

\subsection{Polynomials}

Now we want to verify that, under the expected hypotheses,
the formal power series on the
right hand side of \jimero\ is a polynomial in $\epsilon$ and the
formal variables, modulo exponentially small terms.  (A sharper bound
than we will obtain on the degree of these polynomials is expected.)

The basic fact that we will use is that for $w$ real and non-integral,
$P$ an arbitrary polynomial, and $a$ real,
$$\sum_{n\in \IZ}\exp\left(-\epsilon(n^2+an)+2\pi i wn\right)\cdot P(n)
\eqn\moco$$
vanishes exponentially for $\epsilon\to 0$.  This can be proved,
for instance, by Poisson summation, as in the derivation of \icoco.
There is also an obvious higher dimensional generalization of \moco.
Replace $\IZ$ by a lattice $\Lambda\subset \IR^n$, replace $wn$
by a real-valued linear form $w(n)$ on $\Lambda$ that is not integer-valued,
replace $n^2$ by $(n,n)$, with $(~,~)$ a positive definite quadratic
form on $\IR^n$, and replace $an$ by $(a,n)$ with $a\in\IR^n$.
Then \moco\ becomes
$$\sum_{n\in\Lambda}{\exp\left(-\epsilon\left((n,n)+(a,n)\right)
+2\pi i w(n)\right)
\cdot P(n)}. \eqn\kkkk$$
That this
vanishes exponentially for $\epsilon\to 0$ is proved the same way.

Now, let $\Lambda\subset \IR^n$ be a lattice with an integer-valued
non-degenerate
quadratic form $(~,~)$.  Let $e_i,\,\,\,i=1,\dots ,t$ be not necessarily
distinct non-zero points in $\Lambda$.  Let $H_i$ be the sublattice of
$\Lambda$ defined by $(e_i,h)=0$.  Let us call a sublattice $\Lambda_0
\subset \Lambda$ distinguished if it is non-zero and can be written
as an intersection of a subset of $H_i$'s.  Let $w$ be a real-valued linear
form on $\Lambda$ which is not integer-valued when restricted to any
distinguished $\Lambda_0$.  Let $P$ be a polynomial on $\Lambda$ of
degree $s>t$.  Consider the sum
$$\sum_{h\in\Lambda'}{\exp\left(-\epsilon(h,h)+2\pi i w(h)\right)\cdot P(h)
\over \prod_{j=1}^t (e_j,h) }, \eqn\bxbxbx$$
where $\Lambda'=\Lambda-\cup_jH_j$.  I claim this sum vanishes exponentially
for $\epsilon\to 0$.

Obviously, we can assume that $P$ is a monomial, and thus a product of
linear factors $P_1,\dots,P_s$.    If the $e_j$ are linearly
independent, then we can expand $P_1(h)=\sum_{j=1}^tc_j(e_j,h)$ with
coefficients $c_j$.  It is therefore enough in this case to show that
\bxbxbx\ vanishes exponentially if $P$ is replaced by $P'=
(e_k,h)\cdot P_2\cdot\dots\cdot P_s$.  The factor of $(e_k,h)$ can be
canceled with one factor in the denominator.

At this point, if $e_k$ no longer appears in the list of remaining $e$'s
(recall that the $e$'s may not be distinct), we want to extend the sum in
\bxbxbx\ to run over $\Lambda''=\Lambda-\cup_{j\not= k}H_j$.  By induction
on the dimension of $\Lambda$, we can assume that
$$\sum_{h\in H_k'}{\exp\left(-\epsilon(h,h)+2\pi i w(h)\right)\cdot
P(h)\over \prod_{j\not= k}(e_j,h) } \eqn\xxvvb$$
vanishes exponentially for $\epsilon\to 0$, if $H_k'=H_k
-\cup_{j\not= k}(H_j\cap H_k)$.  As $\Lambda''=\Lambda'\cup H_k'$,
we can replace the sum over $\Lambda'$ by a sum over $\Lambda''$.

After repeating this process, one reduces to the following situation.
The remaining $e_j, \,\,\,j=1\dots t'$ (if any)
are not linearly independent,
and the equations $(e_j,h)=0,\,\,\,j=1\dots t'$ define a non-zero
lattice $\Lambda_1$.  $P$ is a polynomial on this lattice, of degree
$s'>t'\geq 0$.  By the hypothesis about $w$, $w$ is not integer-valued
when restricted to $\Lambda_1$.  The $e_j$ can be regarded as linear
forms on a complementary lattice $\Lambda_2$ to $\Lambda_1$.  The same
Poisson summation used in proving
\kkkk, when applied to the sum over $\Lambda_1$,
now shows that \bxbxbx\ vanishes exponentially for
$\epsilon\to 0$.

To apply this to our problem, let $\Lambda$ be the root lattice of the
compact, connected, and simply-connected Lie group $H$.  We recall from
the theory of compact Lie groups that the dimension of
a representation of highest
weight $h$ is $d(h)=\prod_i(e_i,h+\delta)$,
where $e_i,\,\,\,i=1\dots q$ are the positive roots of $H$,
$\delta $ is one-half their sum, and $(~,~)$ is the usual metric on
$\Lambda$.  To show that \jimero\ is a polynomial in $\epsilon$ and the
formal variables modulo exponentially small terms, it suffices to show
that
$$\sum_h{\lambda_h(u^{-1})\exp\left(-\epsilon(h+\delta,h+\delta)\right)
\cdot P(h+\delta)\over\prod_i(e_i,h+\delta)^{2g-2}} \eqn\moxie$$
vanishes exponentially for $\epsilon\to 0$, if $P(h+\delta)$ is a
Weyl-invariant polynomial of degree greater than $(2g-2)q$, and the sum
runs over dominant weights $h$.  Let $H_i$ be the sublattice of
$\Lambda$ defined by $(e_i,h+\delta)=0$.  Changing variables from
$h$ to $h'=h+\delta$, and using the Weyl-invariance of the numerator
and denominator in \moxie,
we can replace this sum by
$${1\over \#W(H)}\sum_{h'\in\Lambda'}{\lambda_h(u^{-1})\exp\left(-\epsilon
(h',h')\right)\cdot P(h')\over \prod_i(e_i,h')^{2g-2}},\eqn\oxie$$
where $\Lambda'=\Lambda-\cup_i H_i$,
and $W(H)$ is the Weyl group.

In \oxie, $\lambda_h(u^{-1})$ can be written $\exp(2\pi i w(h))$,
where $w$ is some linear form on $\Lambda$.  To deduce the desired property
of \oxie\ from our earlier discussion, it suffices to show that $w(h)$ is
not integer-valued when restricted to any sublattice $\Lambda_0\subset
\Lambda$ defined by vanishing of a subset of the linear forms $(e_i,h')$.

At this point, we specialize to the case $H=SU(N)$ and $u$ a generator of
the center of $H$.  The root lattice
$\Lambda$ can conveniently be taken to consist of $N$-tuples
$(h_1,\dots,h_{N})$, with $\sum_\alpha h_\alpha=0$, and
$h_\alpha-h_\beta\in\IZ$.  The Weyl group is the group of permutations of
the $h$'s.
The linear form $w$ can be written as
$w(h)=kh_1$, with $k$ an integer depending on $u$.
The hypothesis that $u$ generates the center of $SU(N)$ is equivalent to
$(k,N)=1$.  The linear forms $(e_i,h)$ are $h_\alpha-h_\beta$,
$1\leq \alpha<\beta\leq N$.  Any distinguished sublattice
$\Lambda_0$ defined by equations $h_{\alpha_r}-h_{\beta_r}=0$, $r=1\dots m$,
contains a sublattice $\Lambda_0'$ equivalent (up to a Weyl transformation)
to $h_1=\dots =h_m$, $h_{m+1}=\dots =h_N$, with $0<m<N$.
This lattice
contains the point $h_1=\dots =h_m=(N-m)/N$, $h_{m+1}=\dots =
h_N=-m/N$.  If $(k,N)=1$, then $w$ is not an integer at this point,
completing the proof.

This argument shows that \jimero\ is a polynomial in $\epsilon$ and the formal
variables in the expected cases, but (except in the special case
$S^\rho=0$, $Q(\phi)=\Tr\phi^2/8\pi^2)$, the bound on the degree of the
polynomial so obtained is weaker than expected by dimension counting.
It would
be interesting to know how to obtain directly a sharper bound.

\appendix

This appendix is devoted to a simple illustration of
the various localization formulas.  We will consider a situation
with $G=U(1)$ so that we can illustrate both the abelian DH formula and
the new not necessarily abelian formula.  (However, in the abelian
case, the ``new'' formula is a consequence of the DH formula.)
We endow the Lie algebra of $G$
with the
standard metric such that the volume of the group is $2\pi$.

Let $X$ be the two sphere
$$x^2+y^2+z^2 = 1 \eqn\twos$$
and introduce the usual polar coordinates by $z=\cos\theta$,
$x=\sin\theta\,\cos\psi$, $y=\sin\theta\,\sin\psi$.
Thus $0\leq\theta\leq \pi$ and $0\leq\psi\leq 2\pi$.
The usual symplectic volume form is $\omega= \d\cos\theta\,\,\d\psi$.
We consider the $U(1)$ action $\psi\to\psi+{\rm constant}$.
The moment map is
$$\mu = \cos \theta + a,\eqn\turip$$
where $a$ is  an arbitrary constant.
The DH formula therefore applies to
$$Z=\int_{X}\omega\,\,e^{-\beta(\cos\theta+a)}. \eqn\urip$$
Evaluating this explicitly, we have
$$\eqalign{Z=&
\int_{-1}^1\d\cos\theta\int_0^{2\pi}\d\psi e^{-\beta(\cos\theta+a)}
\cr
&={2\pi\over \beta}\left(e^{\beta (1-a)}-e^{-\beta(1+ a)}
\right).\cr}\eqn\orip$$
The two terms can be identified with the contributions from the critical
points $P_\pm$
of $\mu$ at $\cos\theta=\pm 1$.  The factors of $e^{-\beta(a\pm
1)}$
are $e^{-\beta\mu(P_\pm)}$,
while the one loop determinants expanding
around the $P_\pm$ give $\mp 2\pi/\beta$.
A minus sign arises at $P_+$ because it is
a local maximum of $\mu$, unstable in two directions,
each of which, heuristically, contributes a factor of $i$.

Now we want to illustrate the localization formula used
in this paper.  In so doing, we will assume $|a|<1$, leaving the
other (similar) cases to the reader.  Using the formalism of \S2,
we consider the equivariantly closed form
$$\alpha =\exp(\omega+i\phi\mu) \eqn\roff$$
where $\phi$ is a linear function on the one dimensional Lie algebra
of $G$.
Then
$$\oint_X\alpha={1\over {\rm vol}(U(1))}\int_{-\infty}^\infty
{\d\phi\over 2\pi}\int_X \exp\left({\omega+i\phi(\cos\theta+a)}
{-{\epsilon\over 2}\phi^2}\right).  \eqn\picko$$
Integrating over $\psi$ and $\phi$ and setting $x=\cos\theta$,
and ${\rm vol}(U(1))=2\pi$,
we get
$$\oint_X\alpha =\int_{-1}^1{\d x\over (2\pi\epsilon)^{1/2}}
\exp\left(-(x+a)^2/2\epsilon\right).  \eqn\icko$$
The integral would be 1 if the limits were extended from $-\infty$
to $+\infty$.  As it is,
$$\oint_X\alpha= 1-I_+-I_-, \eqn\cko$$
with
$$\eqalign{
I_+ & =
\int_1^\infty {dx\over (2\pi\epsilon)^{1/2}}
\exp\left(-(x+a)^2/2\epsilon\right) \cr
I_- & =
\int_{-\infty}^{-1}{dx\over (2\pi\epsilon)^{1/2}}
\exp\left(-(x+a)^2/2\epsilon\right).  \cr}  \eqn\ixox$$

The three terms correspond to the expected contributions from the critical
points of $I=(\mu,\mu)$.  The absolute minimum of $I$ at $\cos\theta=-a$
can be locally modeled on $T^*G$ and hence
contributes $+1$, as predicted from our general analysis in \S2.3.
The other critical points of $I$ are the critical points $P_\pm$
of $\mu$ that already entered above.
Their contributions are
$-I_\pm$, which are transcendental (error) functions.
The minus signs reflect the fact that the two points $P_{\pm}$ are both local
maxima of $I$.  The contributions of $P_{\pm}$ are
asymptotic for $\epsilon\to 0$  to $\exp(-I(P_\pm)/2\epsilon)$,
in agreement with the general theory.  The complicated error functions
contrast with the elementary functions that
appear as local contributions in the DH formula.  They arise because
although $P_{\pm}$ are nondegenerate critical points of $I$, they
are degenerate critical points of the function $(V,\lambda)^2$
that appears in the key formula \urmor.  By examining \urmor, one can
make quite explicit the fact that the error functions are entirely
determined by the behavior of $\mu$ near $P_{\pm}$ up to second order.

\ack{S. Axelrod made prescient comments several years ago about
two dimensional Yang-Mills theory.
I also want to thank V. Guillemin, C. Vafa, and L. Jeffrey for helpful
comments.}

\refout
\figout
\end